\documentstyle[12pt,amssymb,setspace,graphics]{article}

\input epsf

\begin{document}

\bibliographystyle{apsrev}
\def\half{{1\over 2}}
\def \D {\mbox{D}}
\def\curl {\mbox{curl}\,}
\def \ep {\varepsilon}
\def \lleq {\lower0.9ex\hbox{ $\buildrel < \over \sim$} ~}
\def \ggeq {\lower0.9ex\hbox{ $\buildrel > \over \sim$} ~}
\def\beq{\begin{equation}}
\def\eeq{\end{equation}}
\def\ber{\begin{eqnarray}}
\def\eer{\end{eqnarray}}
\def \apl {ApJ, }
\def \aps {ApJS, }
\def \pd {Phys. Rev. D, }
\def \prl {Phys. Rev. Lett., }
\def \pl {Phys. Lett., }
\def \np {Nucl. Phys., }
\def \l {\Lambda}

\def\apj{{Astroph.\@ J.\ }}
\def\mn{{Mon.\@ Not.\@ Roy.\@ Ast.\@ Soc.\ }}
\def\asta{{Astron.\@ Astrophys.\ }}
\def\aj{{Astron.\@ J.\ }}
\def\prl{{Phys.\@ Rev.\@ Lett.\ }}
\def\pd{{Phys.\@ Rev.\@ D\ }}
\def\nucp{{Nucl.\@ Phys.\ }}
\def\nat{{Nature\ }}
\def\plb {{Phys.\@ Lett.\@ B\ }}
\def \jetpl {JETP Lett.\ }

\begin{center}
\Large{\bf Cosmology and Cosmogony in a Cyclic Universe}
\end{center}
\bigskip

\begin{center}
Jayant V. Narlikar$^{i}$, Geoffrey Burbidge$^{ii}$, R.G. Vishwakarma$^{iii}$
\end{center}

\medskip
\begin{center}
Inter-University Centre for Astronomy and Astrophysics, Pune
411007, India$^{i}$ \\ 
Center for Astrophysics and Space Sciences, University of
California, San Diego, CA 92093-0424, USA$^{ii}$ \\ 
Department of Mathematics, Autonomous University of  Zacatecas, Zacatecas, ZAC C.P. 98060, Mexico$^{iii}$
\end{center}



\begin{abstract}In this paper we discuss the properties of the quasi-steady state
cosmological model (QSSC) developed in 1993 in its role as a
cyclic model of the universe driven by a negative energy scalar
field.  We discuss the origin of such a scalar field in the
primary creation process first described by F. Hoyle and J. V.
Narlikar forty years ago.  It is shown that the creation processes
which takes place in the nuclei of galaxies are closely linked to
the high energy and explosive phenomena, which are commonly
observed in galaxies at all redshifts.

The cyclic nature of the universe provides a natural link between
the places of origin of the microwave background radiation
(arising in hydrogen burning in stars), and the origin of
the lightest nuclei (H, D, He$^3$ and He$^4$).  It also allows us
to relate the large scale cyclic properties of the universe to
events taking place in the nuclei of galaxies.  Observational
evidence shows that ejection of matter and energy from these
centers in the form of compact objects, gas and relativistic
particles is responsible for the population of quasi-stellar
objects (QSOs) and gamma-ray burst sources in the universe.

In the later parts of the paper we briefly discuss the major
unsolved problems of this integrated cosmological and cosmogonical
scheme.  These are the understanding of the origin of the
intrinsic redshifts, and the periodicities in the redshift
distribution of the QSOs.

\end{abstract}



\section{Introduction}
\subsection{Cosmological models}

The standard cosmological model accepted by the majority at
present is centered about the big bang which involves the creation
of matter and energy in an initial explosion.  Since we have
overwhelming evidence that the universe is expanding, the only
alternative to this picture appears to be the classical
steady-state cosmology, of Bondi, Gold and Hoyle, (Bondi and Gold, 1948, 
Hoyle, 1948) or a model in
which the universe is cyclic with an oscillation period which can
be estimated from observation.  In this latter class of model the
bounce at a finite minimum of the scale factor is produced by a
negative energy scalar field.  Long ago Hoyle and Narlikar (1964)
emphasized the fact that such a scalar field will produce models
which oscillate between finite ranges of scale.  In the 1960s
theoretical physicists shied away from scalar fields, and more so
those involving negative energy.  Later Narlikar and Padmanabhan
(1985) discussed how the scalar creation field helps resolve the
problems of singularity, flatness and horizon in cosmology.  It
now appears that the popularity of inflation and the so-called new
physics of the 1980s have changed the 1960s' mind-set.  Thus
Steinhardt and Turok (2002) introduced a negative potential energy
field and used it to cause a bounce from a non-singular high
density state.  It is unfortunate that they did not cite the
earlier work of Hoyle and Narlikar which had pioneered the concept
of non-singular bounce through the agency of a negative energy
field, at a time when the physics community was hostile to these
ideas. Such a field is required to ensure that matter creation
does not violate the conservation of matter and energy.   

Following the discovery of the expansion of the universe by Hubble
in 1929, practically all of the theoretical models considered were
of the Friedmann type, until the proposal by Bondi, Gold and Hoyle
in 1948 of the classical steady state model which first invoked
the creation of matter.  A classical test of this model lay in the
fact that, as distinct from all of the big bang models, it
predicted that the universe must be accelerating (cf Hoyle and
Sandage, 1956). For many years it was claimed that the
observations indicated that the universe is decelerating, and that
this finding disproved the steady state model.  Not until much
later was it conceded that it was really not possible to determine
the deceleration parameter by the classical methods then being used.  Gunn and
Oke (1975) were the first to highlight the observational
uncertainties associated with this test. Of course many other
arguments were used against the classical steady state model (for
a discussion of the history see Hoyle, Burbidge and Narlikar 2000
Chapters 7 and 8). But starting in 1998 studies of the
redshift-apparent magnitude relation for supernovae of Type 1A
showed that the universe $is$ apparently accelerating (Riess et
al. 1998, Perlmutter et al. 1999).  The normal and indeed the
proper way to proceed after this result was obtained should have
been at least to acknowledge that, despite the difficulties associated with
the steady state model, this model had all along been advocating an 
accelerating universe.  

It is worth mentioning that McCrea (1951) was the first to introduce 
vacuum related stresses with equation of state $p = -\rho$ in the context 
of the steady state theory.  Later Gliner (1970) discussed how vacuum-like
state of the medium can serve as original (non singular) state of a Friedmann
model.  

The introduction of dark energy is typical of the way
the standard cosmology has developed; viz,  a new
assumption is introduced specifically to sustain the model against
some new observation.  Thus, when the amount of dark matter proved
to be too high to sustain the primordial origin of deuterium, the
assumption was introduced that most of the dark matter has to be
non-baryonic.  Further assumptions about this dark matter became
necessary, e.g., cold, hot, warm, to sustain the structure
formation scenarios.  The assumption of inflation was introduced
to get rid of the horizon and flatness problems and to do away
with an embarrassingly high density of relic magnetic monopoles. As far
as the dark energy is concerned, until 1998 the general attitude
towards the cosmological constant was typically as summarized by
Longair in the Beijing cosmology symposium: ``None of the
observations to date require the cosmological constant" (Longair
1987).  Yet, when the supernovae observations could not be fitted
without this constant, it came back with a vengeance as dark
energy. 

Although the popularity of the cosmological constant and dark energy
picked up in the late 1990s, there had been earlier attempts at extending
the Friedmann models to include effects of vacuum energy.  A review of
these models, vis-a-vis observations may be found in the article by 
Carroll and Press (1992).  

We concede that with the assumptions of dark energy, non-baryonic
dark matter, inflation etc. an overall self consistent picture has been 
provided within the framework of the standard model.  One demonstration
of this convergence to self consistency is seen from a comparison of a 
review of the values of cosmological parameters of the standard model
by Bagla, et al. (1996), with the present values. Except for the evidence 
from high redshift supernovae, in favour of an accelerating universe which
came 2-3 years later than the above review, there is an overall consistency 
of the picture within the last decade or so, including a firmer belief in the
flat ($\Omega = 1$) model with narrower error bars.  

Nevertheless we also like to emphasize that the inputs required in fundamental 
physics through these assumptions have so far no experimental checks
from laboratory physics.  Moreover an epoch dependent scenario providing
self-consistency checks, e.g. CMB anisotropies, cluster baryon fraction
as a function of redshift does not meet the criterion of `repeatability of 
scientific experiment'.  We contrast this situation with that in stellar
evolution where stars of different masses constitute repeated experimental
checks on the theoretical stellar models thus improving their credibility.  

Given the speculative nature of our understanding of the universe, a sceptic of
the standard model is justified in exploring an alternative avenue wherein
the observed features of the universe are explained with fewer speculative
assumptions.  We review here the progress of such an alternative model.    
  
In this model creation of matter is brought in as a physical phenomenon and a 
negative kinetic energy scalar field is required to ensure that it does  not violate  
the law of conservation of matter and energy. A
simple approach  based on Mach's principle leads naturally to such a field 
within the curved spacetime of general relativity described briefly in $\S$ 2.
The resulting field equations have the two simplest types of solutions for a 
homogeneous and isotropic universe:  (i) those in  which the  universe
oscillates but there is no creation of matter,  and (ii) those in
which the universe steadily expands with a constant value  of
$H_o$ being driven by continuous creation of  matter.  The
simplest model including features of both these solutions is  the
{\it Quasi-Steady  State Cosmology} (QSSC), first proposed  by
Hoyle, Burbidge and Narlikar (1993). It has the scale factor in
the form:

\begin{equation}
S (t) = \exp \bigg({t\over P}\bigg) \{1 + \eta \cos \theta (t)\}, ~\theta (t) \approx \frac{2 \pi t}{Q},
\end{equation}

\noindent where $P$ is the long term `steady state' time scale of expansion
while $Q$ is the period of a single oscillation.

Note that it is essential for the universe to have a long term expansion; for a universe
that has only oscillations without long term expansion would run into problems like
the Olbers paradox.  It is also a challenge in such a model to avoid running into
`heat death' through a steady increase of entropy from one cycle to next. These difficulties
are avoided if there is creation of new matter
at the start of each oscillation as happens in the QSSC, and also, if the universe has a
steady long term expansion in addition to the oscillations.  New matter in such a case
is of low entropy and the event horizon ensures a constant entropy within as the
universe expands.

The  QSSC  has an additional attractive feature if one  uses  the
criterion   of  the  Wheeler  and  Feynman  absorber  theory   of
electromagnetic   radiation  (Wheeler  and  Feynman, 1945, 1949).
This theory provided a very natural explanation of why in
actuality the electromagnetic signals propagate into the future,
i.e., via retarded solutions, despite the time-symmetry of the
basic equations.  By writing the theory in a relativistically
invariant action-at-a-distance form, Wheeler and Feynman showed
that suitable absorptive properties of the universe can lead to
the breaking of time-symmetry.  As was discussed by Hogarth (1962)
and later by Hoyle and Narlikar (1963, 1969, 1971) who also
extended the argument to quantum  electrodynamics,  the
Wheeler-Feynman theory gives results consistent with observations
only if the past absorber is imperfect and the future absorber is
perfect.   This requirement is $not$ satisfied by a  simply cyclic
universe or by an ever-expanding big bang universe but is
satisfied by the QSSC because of expansion  being coupled with
cyclicity.

One may question as to why one needs to have the Wheeler-Feynman
approach to electrodynamics in preference to field theory.  The
advantages are many, including (i) a satisfactory explanation of
the Dirac formula of radiative reaction, (ii) the unambiguous
deduction of why one uses retarded solutions in preference to
advanced ones and (iii) a resolution of the ultraviolet
divergences in quantum electrodynamics.  Rather than go into these
aspects in detail we refer the reader to a recent review by Hoyle
and Narlikar (1995).

Since cosmology seeks to deal with the large-scale properties of
the universe, it inevitably requires a strong connection with
fundamental physics.  In the big bang cosmology particle physics
at very high energy is considered very relevant towards
understanding cosmology.  In the same spirit we believe that the
action at a distance approach to fundamental physics brings about
an intimate link of microphysics with cosmology.  The
Wheeler-Feynman approach is an excellent demonstration of such a
connection.

\subsection{Cosmogony}

In this paper we shall discuss this cosmological model, but first
we want to indicate the importance of the observed behavior of the
galaxies (the observed cosmogony) in this approach.

Now that theoretical cosmologists have begun to look with favor on
the concepts of scalar negative energy fields, and the creation
process, they have taken the position that this subject can only
be investigated by working out models based on  classical
approaches of high energy physics and their effects on the global
scale.  In all of the discussions of what is called precision
cosmology there is no discussion of the remarkable phenomena which
have been found in the comparatively nearby universe showing that
galaxies themselves can eject what may become, new galaxies. We
believe that only when we really understand how individual
galaxies and clusters etc. have formed, evolve, and die (if they
ever do) shall we really understand the overall cosmology of the
universe.  As was mentioned earlier, the method currently used in
the standard model is to suppose that initial quantum fluctuations
were present at an unobservable epoch in the early universe, and
then try to mimic the building of galaxies using numerical
methods, invoking the dominance of non-baryonic matter and dark energy for which
there is no independent evidence.

In one sense we believe that the deficiency of the current
standard approach is already obvious.  The model is based on only
some parts of the observational data.  These are:  all of the
details of the microwave background, the abundances of the light
elements, the observed dimming of distant supernovae, and the large scale
distribution of the observed galaxies.  This has led to the
conclusion that most of the mass-energy making up the universe has
properties which are completely unknown to physics.  This is
hardly a rational position, since it depends heavily on the belief
that all of the laws of physics known to us today can be
extrapolated back to scales and epochs where nothing is really
testable; and that there is nothing new to be learned.

In spite of this, a very persuasive case has been made that all of
the observational parameters can be fitted together to develop
what is now becoming widely accepted as a new standard model, the
so-called $\Lambda$CDM model (Spergel et al., 2003).   There have
been some publications casting doubt on this model, particularly as far as 
the reality of dark energy and cold, dark
matter are concerned (Meyers et al. 2004; Blanchard et al. 2003).  It is
usual to dismiss them as controversial and to argue that a few 
dissenting ideas on the periphery of a generally accepted paradigm 
are but  natural.  However,  it is unfortunately the case that a large fraction of 
our understanding of the extragalactic universe is being based on the
belief that there \textit{was} a beginning and an inflationary phase, and that the 
seeds of galaxies all originate from that very early phase.

We believe that an alternative approach should be considered and
tested by observers and theorists alike. In this scheme the major
themes are (1) that the universe is cyclic and there was no
initial big bang, and (2) \textit{all} of the observational
evidence should be used to test the model.  As we shall show, this
not only includes the observations which are used in the current
standard model, but also the properties and interactions of
galaxies and QSOs which are present in the local ($z < 0.1$)
universe.

Possibly the most perceptive astronomer in recent history was
Viktor Ambartsumian the famous Armenian theorist.  Starting in the
1950s and 1960s (Ambartsumian, 1965) he stressed the role of
explosions in the universe arguing that the associations of
galaxies (groups, clusters, etc.) showed a tendency to expand with
far larger kinetic energy than is expected by assuming that the
gravitational virial condition holds.  

We shall discuss the implications of the cluster dynamics in
Section 6. Here we take up the issue emphasized by Ambartsumian
that there apparently exist phenomena in nuclei of galaxies where
matter seems to appear with large kinetic energy of motion
directed outwards. In Section 6 we will also include other
phenomena that share the same property, namely explosive creation
of matter and energy.  We shall refer to such events as
mini-creation events.

Since these phenomena appear on the extragalactic scale and
involve quasi-stellar objects, active galaxies, powerful radio
sources and clusters and groups of galaxies at all redshifts, we
believe they must have an intimate connection with cosmology.
Indeed, if one looks at standard cosmology, there too the paradigm
centers around the `big bang' which is itself an explosive
creation of matter and energy. In the big bang scenario the origin
of all of the phenomena is ultimately attributed to a single
origin in the very early universe.  No connection has been
considered by the standard cosmologists between this primordial
event and the {\it mini-creation events} (MCEs, hereafter) that
Ambartsumian talked about.  In fact, the QSOs and AGN are commonly
ascribed to supermassive black holes as `prime movers'. In this
interpretation the only connection with cosmology is that it must
be argued that the central black holes are a result of the
processes of galaxy formation in the early universe.

In  the QSSC we have been trying to relate such mini-creation
events (MCEs) directly  to the  large  scale  dynamics of the
universe.  We  show in Sections 2 - 4 that  the dynamics  of
the universe is governed by the frequency and  power of  the
MCEs, and there is a two-way feedback between  the  two. That is,
the universe expands when there is a large MCE  activity and
contracts when the activity is switched off.  Likewise,  the MCE
activity  is  large  when the density  of  the  universe  is
relatively  large and negligible when the density  is  relatively
small.   In  short,  the universe oscillates  between  states  of
finite maximum and minimum densities as do the creation phases in
the MCEs.

This was the model proposed by Hoyle, Burbidge and Narlikar (1993)
and called the {\it quasi-steady state cosmology} (QSSC in brief).
The model was motivated partly by Ambartsumian's ideas and partly
by the growing number of explosive phenomena that are being
discovered in extragalactic astronomy.  In the following sections
we discuss the cosmological model and then turn to the various
phenomena which are beginning to help us understand the basic
cosmogony.  Then we discuss and look at the phenomena themselves
in the framework of this cosmology.  Finally, we discuss some of
the basic problems that have been uncovered by the new
observations for which no theoretical explanation has so far been proposed.

\section{Gravitational Equations With Creation Of Matter}

The mathematical framework for our cosmological model has been
discussed by Hoyle, Burbidge and Narlikar (1995; HBN hereafter), and we outline
briefly its salient features.  To begin with, it is a theory that
is derived from an action principle based on Mach's Principle, and
assumes that the inertia of matter owes its origin to other matter
in the universe. This leads to a theoretical framework wider than
general relativity as it includes terms relating to inertia and
creation of matter. These are explained in the Appendix, and we
use the results derived there in the following
discussion.   

Thus the equations of general relativity are replaced in the
theory by

\begin{equation}
R_{ik} - \frac{1}{2} g_{ik} R + \lambda g_{ik} = 8 \pi
G\bigg[T_{ik} - f \bigg(C_{i} C_{k}
-\frac{1}{4}g_{ik}C^{l}C_{l}\bigg)\bigg],
\end{equation}

\noindent with the coupling constant $f$ defined as

\begin{equation}
f=\frac{2}{3\tau^2}
\end{equation}

\noindent [We have taken the speed of light $c=1.$]  Here
$\tau=\hbar/m_P$ is the characteristic life time of a Planck
particle with mass $m_{P}=\sqrt{3\hbar/8\pi G}$. The gradient of
$C$ with respect to spacetime coordinates $x^{i}(i=0,1,2,3)$ is
denoted by $C_i$.  Although the above equation defines $f$ in
terms of the fundamental constants it is convenient to keep its
identity on the right hand side of Einstein's equations since
there we can compare the $C$-field energy tensor directly with the
matter tensor.  Note that because of positive $f$, the $C$-field
has $negative$ kinetic energy.  Also, as pointed out in the Appendix,
the constant $\lambda$ is $negative$ in this theory.

The question now arises of why astrophysical observation suggests
that the creation of matter occurs in some places but not in
others. For creation to occur at the points $A_0, B_0, \ldots$ it
is necessary classically that the action should not change (i.e. it should 
remain stationary) with respect to small changes in the spacetime
positions of these points, which can be shown to require

\begin{equation}
C_i (A_0) C^i (A_0) = C_i (B_0) C^i (B_0) = \ldots = m_P^2.
\end{equation}

This is in general not the  case:  in general  the  magnitude of
$C_i(X)C^i(X)$ is much  less  that  $m^2_P$. However, as one
approaches closer and closer to the surface of  a massive  compact
body $C_i (X) C^i (X)$ is increased  by  a  general relativistic
time dilatation factor, whereas $m_P$ stays fixed.   

This suggests that we should look for regions of strong gravitational 
field such as those near collapsed massive objects.  In general 
relativistic astrophysics such objects are none other than black holes,
formed from gravitational collapse.  Theorems by Penrose, Hawking
and others (see Hawking and Ellis 1973) have shown that provided
certain positive energy conditions are met, a compact object undergoes
gravitational collapse to a spacetime singularity.  Such objects become
black holes before the singularity is reached.  However, in the present
case, the negative energy of the $C$-field intervenes in such a way 
as to violate the above energy conditions.  What happens to such a
collapsing object containing a $C$-field apart from ordinary matter?  We
argue that such an object does not become a black hole.  Instead,
the collapse of the object is halted and the object bounces back, thanks
to the effect of the $C$-field.  We will refer to such an object as a
compact massive object (CMO) or a near-black hole (NBH).  In the 
following section we discuss the problem of gravitational collapse 
of a dust ball with and without the $C$-field to illustrate this difference.


\section{Gravitational collapse and bounce}

Consider how the classical problem of gravitational
collapse is changed under the influence of the negative energy $C$-field.
First we describe the classical problem which was first discussed by B. Datt (1938).  
We write the spacetime metric inside a collapsing dust ball in comoving coordinates 
($t$, $r$, $\theta$, $\phi$) as

\begin{equation}
{\rm d}s^2 = {\rm d}t^2 - a^2(t)\bigg[\frac{{\rm d}r^2}{1-\alpha r^2} + r^2({\rm d}\theta^2 + {\rm sin}^2 \theta {\rm d} \phi^2)\bigg]
\end{equation}

\noindent where $r$,$\theta$,$\phi$ are constant for a typical dust particle and $t$ is
its proper time.  Let the dust ball be limited by $r \leq r_b$.  

In the above problem we may describe the onset of collapse 
at $t=0$ with $a(0) = 1$ and $\dot{a}(0)=0$.  The starting density $\rho_0$ is
related to the constant $\alpha$ by

\begin{equation}
\alpha = \frac{8 \pi G \rho_0}{3}.
\end{equation}

The field equations (2) {\it without} the $C$-field and the cosmological constant 
then tell us that the equation of collapse is given by 

\begin{equation}
\dot{a}^2 = \alpha \bigg(\frac{1-a}{a}\bigg),
\end{equation}

\noindent and the spacetime singularity is attained when $a(t) \rightarrow 0$ as $t \rightarrow t_S$,
where 

\begin{equation}
t_S = \frac{\pi}{2\sqrt{\alpha}}.
\end{equation}

Note that we have ignored the $\lambda$- term as it turns out to have a negligible effect
on objects of size small compared to the characteristic size of the universe.  

The collapsing ball enters the event horizon at a time $t=t_H$ when

\begin{equation}
r_b a(t_H) = 2GM,
\end{equation}

\noindent where the gravitational mass of the dust ball is given by 

\begin{equation}
M = \frac{4 \pi}{3} r^3_b \rho_0 = \frac{\alpha r^3_b}{2G}.
\end{equation}

\noindent This is the stage when the ball becomes a black hole.  

When we introduce an ambient $C$-field into this problem, it gets 
modified as follows.  In the homogeneous situation under discussion,
$C$ is a function of $t$ only.  Let, as before $a(0) = 1$, $\dot{a}(0) = 0$ and 
let $\dot{C}$ at $t=0$, be given by $\beta$. Then it can be easily seen that the equation 
(7) is modified to

\begin{equation}
\dot{a}^2 = \alpha \bigg(\frac{1-a}{a}\bigg) - \gamma \bigg(\frac{1-a}{a^2}\bigg)
\end{equation}

\noindent where $\gamma = 2 \pi Gf \beta^2>0$.  Also the earlier relation (6) is modified
to

\begin{equation}
\alpha = \frac{8 \pi G \rho_0}{3} - \gamma.  
\end{equation}

It is immediately clear that in these modified circumstances $a(t)$ cannot reach
zero, the spacetime singularity is averted and the ball bounces at a minimum value
$a_{min} >0$, of the function $a(t)$.   

Writing $\mu = \gamma / \alpha$, we see that the second zero of $\dot{a}(t)$ 
occurs at $a_{{\rm min}} = \mu$.  Thus even for an initially weak $C$-field, we get a 
bounce at a finite value of $a(t)$.

But what about the development of a black hole?  The gravitational mass of the
black hole at any epoch $t$  is estimated by its energy content, i.e., by, 

\begin{eqnarray}
M(t) & = & \frac{4 \pi}{3} r^3_b a^3(t)\bigg\{\rho - \frac{3}{4}f \dot{C}^2 \bigg\} \nonumber \\
& = & \frac{\alpha r^3_b}{2G}\bigg(1 + \mu - \frac{\mu}{a}\bigg).
\end{eqnarray}

Thus the gravitational mass of the dust ball $decreases$ as it contracts and
consequently its effective  Schwarzschild radius decreases.  This happens 
because of the reservoir of negative energy whose intensity rises faster 
than that of dust density.  Such a result is markedly different from that for 
a collapsing object with positive energy fields only.  From (13) we have the ratio

\begin{equation}
F \equiv \frac{2GM(t)}{r_b a(t)} = \alpha r^2_b \bigg\{\frac{1 + \mu}{a} - \frac{\mu}{a^2}\bigg\}.
\end{equation}

Hence,

\begin{equation}
\frac{dF}{da} = \frac{\alpha r^2_b}{a^2} \bigg\{\frac{2 \mu}{a} - (1 + \mu) \bigg\}.
\end{equation}

We anticipate that $\mu \ll 1$, i.e., the ambient $C$-field energy density is much
less than the initial density of the collapsing ball.  Thus $F$ increases as $a$ decreases
and it reaches its maximum value at $a \cong 2 \mu$.  This value
is attainable, being larger than $a_{{\rm min}}$.  Denoting this with $F_{{\rm max}}$, we get 

\begin{equation}
F_{{\rm max}} \cong \frac{\alpha r^2_b}{4 \mu}.
\end{equation}

In general $\alpha r^2_b \ll 1$ for most astrophysical objects.  For the Sun, $\alpha r^2_b \cong 4 \times 10^{-8}$, while for a white dwarf it is $\sim 4 \times 10^{-6}$.  We assume that 
$\mu$, although small compared to unity, exceeds such values, thus making $F_{max} <1$. 
{\it In such circumstances black holes do not form.}

We consider scenarios in which the object soon after bounce picks up high outward
velocity.  From (11) we see that maximum outward velocity is attained at $a=2 \mu$ and
it is given by 

\begin{equation}
\dot{a}^2_{{\rm max}} \approx \frac{\alpha}{4 \mu}.
\end{equation}

\noindent As $\mu \ll 1$, we expect $\dot{a}_{{\rm max}}$ to attain high values.  Likewise
the $C$-field gradient ($\dot{C}$ in this case) will attain high values in such cases.  

Thus, such objects after bouncing at $a_{min}$ will expand and as $a(t)$ increases the
strength of the $C$-field falls while for small $a(t)$ $\dot{a}$ increases rapidly as per
equation (11).  This expansion therefore resembles an explosion.  Further, the high 
local value of the $C$-field gradient will trigger off creation of Planck particles.   We
will return to this explosive phase in section 7 to illustrate its relevance to high energy 
phenomena. 

It is worth stressing here that even in classical general
relativity, the external observer never lives long enough to
observe the collapsing object enter the horizon.  Thus all claims
to have observed black holes in X-ray sources or galactic nuclei
really establish the existence of compact massive objects, and as
such they are consistent with the NBH concept.  A spinning NBH, 
for example can be approximated by the Kerr solution limited to region 
outside the horizon (- in an NBH there is no horizon).  In cases 
where $\dot{C}$  has not gone to the level of creation of matter, 
an NBH will behave very much like a Kerr black hole.  

The theory would profit most from a quantum description of the
creation process. The difficulty, however, is that
Planck particles are defined as those for which the Compton
wavelength and the gravitational radius are essentially the same,
which means that, unlike other quantum processes, flat spacetime
cannot be used in the formulation of the theory. A gravitational
disturbance is necessarily involved and the ideal location for
triggering creation is that near a CMO. The $C$-field boson far
away from a compact object of mass $M$ may not be energetic enough
to trigger the creation of a Planck particle.  On falling into the
strong gravitational  field of a sufficiently compact object,
however, the boson energy is multiplied by a factor,  $(1 - 2GM /
r) ^{-1/2}$ for a local Schwarzschild metric.
 
Bosons then multiply up in a cascade, one makes two, two makes
four, $\ldots$, as in the discharge of a laser, with particle
production multiplying up similarly and with negative pressure
effects ultimately blowing the system apart. This is the explosive
event that we earlier referred to as a {\it mini-creation event}
(MCE).  Unlike the big bang, however, the dynamics of this
phenomenon is {\it well defined and non-singular}.  For a detailed
discussion of the role of a NBH as well as the mode of its
formation, see Hoyle et al. (2000), (HBN hereafter) p. 244-249.

While still qualitative, we shall show that this view agrees well
with the empirical facts of observational astrophysics. For, as
mentioned in the previous section, we do see several explosive
phenomena in the universe, such as jets from radio sources, gamma
ray bursts, X-ray bursters, QSOs and active galactic nuclei, etc.
Generally it is assumed that a black hole plays the lead role in
such an event by somehow converting a fraction of its huge
gravitational energy into large kinetic energy of the `burst'
kind. In actuality, we do not see infalling matter that is the
signature of a black hole.  Rather one sees outgoing matter and
radiation, which agrees very well with the explosive picture
presented above.
\section{Cosmological Models}

The qualitative picture described above is too difficult and
complex to admit an exact solution of the field equations (2).
The problem is analogous to that in standard cosmology where a
universe with inhomogeneity on the scale of galaxies, clusters,
superclusters, etc., as well as containing dark matter and
radiation is impossible to describe exactly by a general
relativistic solution. In such a case one starts with simplified
approximations as in models of Friedmann and Lemaitre and then
puts in specific details as perturbation.  The two phases of
radiation-dominated and matter-dominated universe likewise reflect
approximations implying that in the early stages the relativistic
particles and photons dominated the expansion of the universe
whereas in the later stages it was the non-relativistic matter or
dust, that played the major role in the dynamics of the universe.

In the same spirit we approach the above cosmology by a
mathematical idealization of a homogeneous and isotropic universe
in which there are regularly phased epochs when the MCEs were
active and matter creation took place while between two
consecutive epochs there was no creation (- the MCEs lying
dormant).  We will refer to these two situations as creative and
non-creative modes.  In the homogeneous universe assumed here 
the $C$-field will be a function of cosmic time only.  We will be interested 
in the matter-dominated analogues of the standard models since, as we shall 
see, the analogue of the radiation-dominated state never arises except
locally in each MCE where, however, it remains less intense than
the $C$-field.  In this approximation, the increase or decrease of
the scale factor $S(t)$ of the universe indicates an average
smoothed out effect of the MCEs as they are turned on or off.  The
following discussion is based on the work of Sachs, et al. (1996).

We  write the field equations (2) for the  Robertson-Walker line
element  with $S(t)$ as scale factor and $k$ as curvature parameter and for matter in 
the form of dust,  when  they reduce to essentially two independent equations~:

\begin{equation}
2 {\ddot S \over S} + {\dot S^2 + k \over S^2} = 3 \lambda + 2 \pi
G f\dot C^2
\end{equation}

\begin{equation}{3 (\dot S^2 + k)\over S^2} =  3 \lambda + 8 \pi G \rho - 6
\pi G f\dot C^2,
\end{equation}

\noindent  where we have set the speed of light $c = 1$ and the
density of dust is given by $\rho$.  From these  equations we get
the conservation law in the form of an identity~:

\begin{equation}
{d \over dS} \{S^3 (3\lambda + 8 \pi G \rho - 6 \pi G f \dot
C^2)\} = 3S^2 \{3\lambda + 2 \pi G f \dot C^2\}.
\end{equation}

\noindent  This law incorporates ``creative'' as well as
``non-creative'' modes. We will discuss both in that order.

\subsection{The creative mode}

\noindent  This has

\begin{equation}
T^{ik}_{~;k} \not= 0
\end{equation}

\noindent  which, in terms of our simplified model becomes

\begin{equation}
{d\over dS} (S^3\rho) \not= 0.
\end{equation}

\noindent  For the case $k = 0$, we get a simple steady-state de
Sitter  type solution with

\begin{equation}
\dot C = m,~~~~~~ S = \exp (t/P),
\end{equation}

\noindent  and from  (18) and (19) we get

\begin{equation}
\rho = f m^2, ~~~~~~ {1\over P^2} = {2\pi G \rho \over 3} +
\lambda.
\end{equation}

\noindent  Since $\lambda < 0$, we expect that

\begin{equation}
\lambda \approx - {2\pi G\rho \over 3},~~~~  {1\over P^2}
 \ll \vert \lambda \vert,
\end{equation}

\noindent  but will defer the determination of $P$ to after we
have looked at the non-creative solutions.  Although Sachs, et al. (1996)
have discussed all cases, we will concentrate on the simplest one of 
flat space $k=0$.  

The rate of creation of matter is given by

\begin{equation}
J = {3\rho \over P}.
\end{equation}

\noindent  As  will be seen in the quasi-steady state case, this
rate of  creation is  an  overall average made of a  large  number
of  small events.  Further, since the creation activity has ups
and  downs, we expect $J$ to denote some sort of temporal average.
This  will become  clearer after we consider the non-creative mode
and  then link it to the creative one.

\subsection{The non-creative mode}

\noindent  In this case $T^{ik}_{~;k} = 0$ and we  get  a
different set of solutions. The conservation of matter alone gives

\begin{equation}
\rho \propto {1\over S^3},
\end{equation}

\noindent while for (27) and a constant $\lambda$, (20) leads to

\begin{equation}
\dot C \propto {1\over S^2}.
\end{equation}

\noindent  Therefore, equation (19) gives

\begin{equation}
{\dot S^2 + k \over S^2} = \lambda + {A \over S^3} - {B \over
S^4},
\end{equation}

\noindent  where  $A$ and $B$ are positive constants  arising from
the constants  of proportionality in (27) and (28). We now find that the exact 
solution of (29) in the case $k = 0$, is given by

\begin{equation}
S = \bar S [1 + \eta \cos \theta (t)]
\end{equation}

\noindent  where $\eta$ is a parameter and the function $\theta
(t)$ is given by

\begin{equation}
\dot \theta^2 = -\lambda (1 + \eta \cos \theta)^{-2} \{6    +
4\eta   \cos   \theta   +   \eta^2   (1    +    \cos^2 \theta)\}.
\end{equation}

\noindent  Here, $\bar S$ is a constant and the parameter
$\eta$ satisfies the condition: $\vert \eta \vert < 1.$
Thus the scale factor never becomes zero and  the model oscillates
between finite scale limits

\begin{equation}
S_{\rm min} \equiv \bar S (1 - \eta) \leq S \leq \bar S (1 + \eta)
\equiv S_{\rm max},
\end{equation}

\noindent  The density of matter and the   $C$-field energy
density are given by

\begin{equation}
\bar \rho = -{3\lambda \over 2\pi G} (1 + \eta^2),
\end{equation}

\begin{equation}
f\dot    C^2   =   -{\lambda   \over   2\pi    G}    (1-\eta^2)
(3+\eta^2),
\end{equation}

\noindent  while the period of oscillation is given by

\begin{equation}
Q = {1\over \sqrt{-\lambda}} \int^{2\pi}_0 {(1+\eta \cos \theta)
d\theta \over
\{6 + 4\eta \cos \theta + \eta^2 (1 + \cos^2 \theta)\}^{1/2}}.
\end{equation}

\noindent The oscillatory solution can be approximated by a
simpler sinusoidal solution with the same period ~:

\begin{equation}
S \approx 1 + \eta {\rm ~cos} \frac{2\pi t}{Q}.
\end{equation}

\noindent Thus the function $\theta(t)$ is approximately
proportional to $t$.

Notice that there is considerable similarity between the
oscillatory solution obtained here and that discussed by
Steinhardt and Turok (2002) in the context of a scalar field
arising from phase transition.  The bounce at finite minimum of
scale factor is produced in both cosmologies through a negative
energy scalar field.  As we pointed out in the introduction, Hoyle
and Narlikar (1964) [{\it see also} Narlikar (1973)] have
emphasized the fact that such a scalar field can produce models
which oscillate between finite ranges of scale.  In the
Hoyle-Narlikar paper cited above $\dot{C}\propto 1/S^3$, as
opposed to (28), exactly as assumed by Steinhardt and Turok (2002)
38 years later. This is because instead of the trace-free energy
tensor of Equation (2) here, Hoyle and Narlikar had used the
standard scalar field tensor given by

\begin{equation}
- f \bigg(C_iC_k - \frac{1}{2} g_{ik} C_lC^l \bigg).
\end{equation}

Far from being dismissed as physically unrealistic, negative
kinetic energy fields like the $C-$field are gaining popularity.
Recent works by Rubano and Seudellaro (2004), Sami and 
Toporensky (2004), Singh, et al. (2003) who refer to the earlier work by
Hoyle and Narlikar (1964) have adapted the same ideas to describe
phantom matter and the cosmological constant. In these works
solutions of vacuum field equations with a cosmological constant
are interpreted as a steady state in which matter or entropy is
being continuously created.  Barrow, et al. (2004) who obtain 
bouncing models similar to ours refer to the paper by Hoyle and 
Narlikar (1963) where $C$-field idea was proposed in the context of
the steady state theory.

\subsection{The Quasi-Steady State Solution}

The  quasi-steady state cosmology is described by a  combination
of the creative and the non-creative modes.  For this the general
procedure to be followed is to look for a composite solution of the form

\begin{equation}
S (t) = \exp \bigg({t\over P}\bigg) \{1 + \eta \cos \theta (t)\}
\end{equation}

\noindent  wherein    $P \gg Q$. Thus over a period $Q$ as given
by (35), the universe is  essentially in  a non-creative mode.
However, at  regular  instances separated  by the period $Q$ it
has injection of new  matter  at such  a  rate  as to preserve an
average rate  of  creation  over period   $P$  as  given by $J$ in
(26). It is  most  likely that these epochs  of creation are those
of the minimum value of  the  scale factor during oscillation when
the level of the $C$-field background is the highest. There is a sharp
drop at a typical minimum but the $S(t)$ is a continuous curve with
a zero derivative at $S = S_{{\rm min}}$.  

Suppose  that  matter creation takes place  at  the  minimum value
of $S = S_{\rm min}$, and that  $N$ particles are  created  per
unit  volume  with  mass $m_0$.  Then the  extra  density  added
at this epoch in the creative mode is

\begin{equation}
\Delta \rho = m_0 N.
\end{equation}

\noindent  After one cycle the volume of the space expands by a
factor $\exp~ (3Q / P)$ and to restore the density to its original
value we should have

\begin{equation}
(\rho + \Delta \rho) e^{-3Q/P}  = \rho,~~~{\rm i.e.},~~ \Delta
\rho / \rho \cong 3Q/P.
\end{equation}

The $C$-field strength likewise takes a jump at creation and
declines over the following cycle by the factor $\exp (-4 Q/P)$.
Thus the  requirement of ``steady state'' from cycle to cycle
tells  us that the change in the strength of $\dot C^2$ must be

\begin{equation}
\Delta \dot C^2 = {4Q \over P} \dot C^2.
\end{equation}

\noindent   The above result is seen to be consistent with (40)
when we take note of the conservation law (20).  A little
manipulation of this equation gives us

\begin{equation}
{3\over 4} {1\over S^4} {d \over dS} (f \dot C^2 S^4) = {1\over
S^3} {d \over dS} (\rho S^3).
\end{equation}

\noindent  However,  the right hand side is the rate of creation
of  matter per unit volume.  Since from (40) and (41) we have

\begin{equation}
{\Delta \dot C^2 \over \dot C^2} = {4\over 3} {\Delta \rho \over
\rho},
\end{equation}

\noindent  and from  (23) and (24)  we have $\rho = f\dot C^2$, we
see  that  (42)  is deducible from (40) and (41).

To  summarize, we find that the composite solution  properly
reflects  the  quasi-steady state character of the  cosmology  in
that  while each cycle of duration $Q$ is exactly a repeat of the
preceding  one, over a long time scale the universe expands with
the de Sitter expansion factor $\exp (t/P)$.  The two time scales
$P$ and $Q$ of the  model thus turn out to be related to the
coupling  constants and  the parameters $\lambda, f, G, \eta$ of
the field  equations.   Further progress  in the theoretical
problem can be made after we  understand the quantum theory of
creation by the $C$-field.

These solutions contain sufficient number of arbitrary  constants
to assure us that they are generic, once we make the
simplification that the universe obeys the Weyl postulate and the
cosmological principle. The composite solution can be seen as an
illustration of how a non-creative mode can be joined with the
creative mode.  More possibilities may exist of combining the  two
within  the given framework. We have, however, followed the
simplicity argument (also used in the standard big bang
cosmology) to  limit our present choice to the composite solution
described here.  HBN have used (38), or its approximation

\begin{equation}
S(t) = {\rm exp} \bigg(\frac{t}{P}\bigg)\bigg\{1+\eta {\rm
~cos~}\frac{2\pi t}{Q}\bigg\}
\end{equation}

\noindent to work out the observable features of the QSSC, which
we shall highlight next.

\section{The Astrophysical Picture}

\subsection{Cosmological Parameters}

Coming next to a physical interpretation of  these mathematical
solutions, we can visualize the above model in terms of the
following values of its parameters:

\begin{eqnarray}
P=20Q,~Q=5~\times 10^{10} {\rm yrs}, ~\eta=0.811, \nonumber \\
\lambda=-0.358~\times 10^{-56} ({\rm cm)}^{-2}.
\end{eqnarray}

\noindent To fix ideas, we have taken the maximum redshift
$z_{max} = 5$ so that the scale factor at the present epoch
$S_{0}$ is determined from the relation $S_{0} = \bar  S
(1 - \eta)(1 + z_{max})$.  This set of parameters has been used in
recent papers on the QSSC (Narlikar, et al. 2002, 2003).  For this
model the ratio of maximum to minimum scale factor in any
oscillation is around 9.6.

These parametric values are not uniquely chosen; they
are rather indicative of the magnitudes that may describe the real
universe.  For example, $z_{max}$ could be as high as 10 without placing 
any strain on the model.  The various observational tests seek to place
constraints on these values.  Can the above model quantified by
the above parameters cope with such tests? If it does we will know
that the QSSC provides a realistic and viable alternative to the
big bang.

\subsection{The Radiation Background}

As far as the origin and nature of the CMBR is concerned we use a
fact that is always ignored by standard cosmologists.  If we
suppose that most of the $^4$He found in our own and external
galaxies (about 24\% of the hydrogen by mass) was synthesized by
hydrogen burning in stars, the energy released amounts to about
4.37 x 10$^{-13}$ erg cm$^{-3}$.  This is almost exactly equal to
the energy density of the microwave background radiation with T =
2.74$^\circ$K. For standard cosmologists this has to be dismissed
as a coincidence, but for us it is a powerful argument in favor of
the hypothesis that the microwave radiation at the level detected
is relic starlight from previous oscillations in the QSSC which
has been thermalized (Hoyle, et al. 1994).  Of course, 
this coincidence loses its significance in the standard big bang 
cosmology where the CMBR temperature is epoch-dependent.  

It is then natural to suppose that the other light isotopes,
namely D, $^3$He, $^6$Li, $^7$Li, $^9$Be, $^{10}$B and $^{11}$B
were produced by stellar processes.  It has been shown (cf.
Burbidge and Hoyle, 1998) that both spallation and stellar
flares (for $^2$D) on the surfaces of stars can explain the
measured abundances. Thus {\it all} of the isotopes are
ultimately a result of stellar nucleosynthesis (Burbidge et al.
1957; Burbidge and Hoyle 1998). 

This option raises a problem, however.  If we simply extrapolate our
understanding of stellar nucleosynthesis, we will find it hard to explain
the relatively low metallicity of stars in our Galaxy. This is still an unsolved problem.  
We believe but have not yet established that it may be that the initial mass function 
of the stars where the elements are made is dominated by stars which are
only able to eject the outer shells while all of the heavy
elements are contained in the cores which simply collapse into black
holes.  Using theory we can construct a mass function which will
lead to the right answer (we think) but it has not yet been done. 
But of course our handwaving in this area is no better than all of
the speculations that are being made in the conventional approach when
it comes to the ``first" stars.

The theory succeeds in relating the intensity and temperature of
CMBR to the stellar burning activity in each cycle, the
result emphasizing the causal relationship between the
radiation background and nuclear abundances. But, how is the
background thermalized?  The metallic whisker shaped grains
condensed  from supernova  ejecta have been shown to effectively
thermalize  the relic   starlight  (Hoyle  et  al., 1994, 2000).
It  has  also   been demonstrated  that inhomogeneities on the
observed  scale  result from  the  thermalized  radiation  from
clusters, groups of galaxies  etc. thermalized at the minimum of 
the last oscillation (Narlikar et al., 2003).
 By using a toy
model for these sources, it has been shown that the resulting angular power
spectrum has a satisfactory fit to the data compiled by Podariu et al (2001)
for the band power spectrum of the CMBR temperature inhomogeneities. Extending
that work further we show, in the following, that the model is also consistent
with the first- and third- year observations of the Wilkinson Microwave 
Anisotropy Probe (WMAP) (Page et al. 2003; Spergel et al. 2006).

Following Narlikar et al (2003) we model the inhomogeneity of the CMBR 
temperature as a
set of small disc-shaped spots, randomly distributed on a unit
sphere. The spots may be either `top hat' type or `Gaussian'
type. In the former case they have sharp boundaries whereas in the
latter case they taper outwards.  We  assume the former for
clusters, and the latter for the galaxies, or groups of galaxies, and also for
the curvature effect.  This is because the clusters will
tend to have rather sharp boundaries whereas in the other cases
such sharp limits do not exist. The resultant inhomogeneity of the CMBR
thus arises from a superposition of random spots of three
characteristic sizes corresponding to the three effects - the curvature effects
at the last minimum of the the scale factor, clusters, and groups of galaxies.
This is given by a $7$ - parameter model of the angular power spectrum
(for more details, see Narlikar et al, 2003):
\begin{eqnarray}
{\mathcal C}_{l} &=& A_1 \,\,~ l(l+1) e^{-l^{2}\alpha^{2}_{1}}\nonumber \\ {}&&  + 
A_2\,\, \frac{l^{\gamma-2}}{l+1}\left[{\rm ~cos~} \alpha_{2} P_{l} ({\rm cos}~ \alpha_{2}) 
- P_{l-1}({\rm cos}~ \alpha_{2})\right]^2 \nonumber \\ {}&& + A_3\,\, l(l+1)e^{-l^{2}\alpha^{2}_{3}},\label{eq:cl}
\end{eqnarray}
where the parameters $A_1$, $A_2$, $A_3$ depend
on the number density as well as the typical temperature fluctuation
of each kind of spot, the parameters $\alpha_{1}$, $\alpha_{2}$,
$\alpha_{3}$ correspond to the multipole value $l_p$ at which the
${\cal C}_l$ from each component peaks, and the parameter $\gamma$ refers
to the correlation of the hot spots due to clusters.
These parameters are determined by fitting the model to the  
observations by following the method we have used in (Narlikar, et al, 2003).
We find that the observations favour a constant  in place of 
the first gaussian profile in equation (\ref{eq:cl}), resulting in a 
 6-parameter model with $A_1$, $A_2$, $A_3$, 
$\alpha_{2}, \alpha_{3}$ and $\gamma$ as the remaining free parameters.
We should mention that the first gaussian profile of equation (\ref{eq:cl}) 
had been 
conjectured by Narlikar, et al (2003) to be related to signature of spacetime 
curvature at the last minimum scale of oscillation. This conjecture was 
analogous to the particle horizon in the standard cosmology. In the QSSC, 
there is no particle horizon and the current observations suggest that the 
curvature effect on CMBR inhomogeneity is negligible.

For the actual fitting, we consider the WMAP-three year data release 
(Spergel, et al, 2006). The data for the mean value of 
TT power spectrum have been binned into 39 bins in multipole space.
We find that the earlier fit (Narlikar, et al, 2003) of the model 
is worsened when we consider the new data, giving $\chi^2 = 129.6$ 
at 33 degrees of freedom. 
However, we should note that while the new data set (WMAP-three year) has 
generally increased its accuracy, compared with the WMAP-one year observations, for $l\leq700$, the observations for higher $l$ do not seem to agree. 
This is clear from Figure 1 where we have shown these two observations 
simultaneously. If we exclude the last three points from the fit, we can
have a satisfactory fit giving 
$\chi^2 = 83.6$ for the best-fitting parameters  
 $A_1=890.439\pm26.270$, $A_2=2402543.93\pm3110688.86$, $A_3=0.123\pm0.033$, 
$\alpha_2 = 0.010\pm0.0001$, $\alpha_3 = 0.004\pm0.000004$ and 
$\gamma = 3.645\pm0.206$,
We shall see in the following that the standard cosmology also supplies a 
similar fit to the data. It should be noted that the
above mentioned parameters in the QSSC can be related to the physical 
dimensions 
of the sources of inhomogeneities along the lines of Narlikar et al (2003)
and are within the broad range of values expected from the 
physics of the processes.

For comparison, we fitted  the same binned data, to the anisotropy spectrum
prediction of a grid of open-CDM and $\Lambda$-CDM models within the
standard big bang cosmology. We varied the matter density, $\Omega_{\rm m}
=0.1$ to $1$ in steps of $0.1$; the baryon density, $\Omega_{\rm b} h^2$
from $0.005$ to $0.03$ in steps of $0.004$ where $h$ is the Hubble
constant in units of $100$ km s$^{-1}$ Mpc$^{-1}$; and the age of the
universe, $t_0$ from $10$ Gyr to $20$ Gyr in steps of $2$ Gyr.  
For each value of $\Omega_{\rm m}$ we considered an open model and
one flat where a compensating $\Omega_\Lambda$ was added. 
For the same binned data set, we find that the minimum value of
$\chi^2$ is obtained for the flat model  
($\Omega_{\rm m}=0.2=1-\Omega_\Lambda$, $\Omega_{\rm b} h^2=0.021$, 
$t_0=14$ Gyr and $h=0.75$) with $\chi^2$=95.9 for the full data and
$\chi^2$=92.7 from the first 36 points. Though the fit can be improved
marginally by fine tunning the parameters further. However, it should be noted
that the error bars (we have used) provided by the WMAP team 
 provide only a rough estimate of the errors, not the
exact error bars. For a proper assignment of errors, it is suggested to
use the complete Fisher matrix. However, one should note that some components 
that go into making 
the Fisher matrix, depend on the particular models. This makes the errors
model dependent which  prohibits an independent assessment of the viability of 
the model. Hence until the model-independent errors are available from the 
observations, we are satisfied by our procedures and qualities of fit for both
theories.
\bigskip

\begin{figure}[tbh!]
\centerline{{\epsfxsize=14cm {\epsfbox[50 250 550 550]{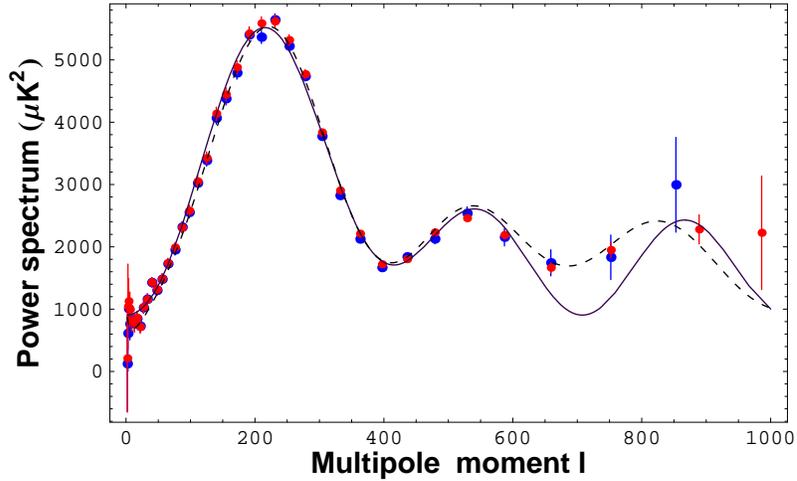}}}}
{\caption{\small We plot the best-fitting angular power spectrum
curves to the WMAP-three year data (shown in red colour) averaged into 39 bins. 
The continuous curve corresponds to the QSSC with 6 parameters and
the dashed one to the big bang model with $\Omega_{\rm m}=0.2$,
$\Omega_\Lambda=0.8$. We notice that the highest parts of contribution to 
$\chi^2$ is from the last three points and the first 4 points of the data, 
on which the 
observations have not settled yet, as is clear from the comparision of these 
data with the WMAP-one year 
data (shown in blue colour). The rest of the points have reasonable fits with 
the theoretical curves.}}
\end{figure}

Figure 1 shows the best-fitting angular power spectrum curve obtained for 
QSSC by using the six parameter model. For comparison, we have also drawn
the best-fitting big bang model.

We mention in passing that recent work (Wickramasinghe 2005) indicates that small
traces of polarization would be expected in the CMBR wherever it passes through 
optically thin clouds of iron whiskers.  These whiskers being partially aligned along
the intracluster magnetic fields will yield a weak signal of polarization on the scale
of clusters or smaller ojects.   

It should be noted that the small scale anisotropies do not constitute 
as crucial a test for our model as they do for standard cosmology.  Our general
belief is that the universe is inhomogeneous on the scales of galaxy-cluster-supercluster
and the QSSC model cannot make detailed claims of how these would result in the
anisotropy of CMBR.  In this respect, the standard model subject to all its assumptions
(dark matter, inflation, dark energy, etc.) makes much more focussed predictions
of CMBR anisotropy.  
 
It is worth commenting on another issue of an astrophysical
nature. The typical QSSC cycle has a lifetime long enough for most
stars of masses exceeding $\sim 0.5-0.7 M_{\odot}$ to have burnt
out. Thus stars from previous cycles will be mostly extinct as
radiators of energy.  Their masses will continue, however, to
exert a gravitational influence on visible matter. The so-called
dark matter seen in the outer reaches of galaxies and within
clusters may very well be made up, at least in part, of these
stellar remnants.

To what extent does this interpretation tally with observations?
Clearly, in the big bang cosmology the time scales are not long
enough to allow such an interpretation.  Nor does that cosmology
permit dark matter to be baryonic to such an extent. The
constraints on baryonic dark matter in standard cosmology come
from (i) the origin and abundance of deuterium and (ii)
considerations of large scale structure.  The latter constraint further 
requires the nonbaryonic matter to be cold. In the QSSC, as has been
shown before, these constraints are not relevant. For other
observational issues completely handled by the QSSC, see Hoyle et
al. (2000). 

The QSSC envisages stars from previous cycles to have burnt out and
remained in and around their parent galaxies as dark matter.  These may 
be very faint white dwarfs, neutron stars and even more massive remnants 
of supernovae, like near black holes.  Their masses may be in the neighbourhood
of $M_\odot$, or more, i.e., much larger than planetary or brown dwarf masses.  Thus 
one form of baryonic dark matter could be in such remnants.  In this connection results 
from surveys like MACHO or OGLE would provide possible constraints on
this hypothesis.  We should mention here that unlike the standard cosmology, 
the QSSC does not have limits on the density of baryonic matter from 
considerations of deuterium production or formation of large scale structure.

\section{Explosive Cosmogony}

\subsection{Groups and clusters of galaxies}

We have already stated that it was Ambartsumian (1965) who first
pointed out that the simplest interpretation of many physical
systems of galaxies ranging from very small groups to large
clusters is that they are expanding and coming apart.  Since most
of the observations are of systems at comparatively small
redshifts it is clear that this takes place at the current epoch,
and while we do not have direct evidence of the situation at large
redshifts, it is most likely a general phenomenon.

Why has this effect been so widely ignored?  The answer to this is
clearly related to the beliefs of earlier generations of
cosmologists.  From an historical point of view, the first
physical clusters were identified in the 1920s, and it was Zwicky,
and later others who supposed that they must be stable systems. By
measuring individual redshifts of a number of the galaxies in such
a cluster it is possible to get a measurement of the line-of-sight
random motions.  For stability the virial condition $2E_K + \Omega
= O$ needs to be satisfied where $E_K$ and $\Omega$ are the
average values of the kinetic energy and potential energy of the
cluster members. Extensive spectroscopic studies from the 1950s
onward showed that nearly always the kinetic energy of the visible
matter far exceeds the potential energy apparent from the visible
parts of the galaxies. Many clusters have structures which suggest
they are stable and relaxed.  Thus it was deduced that in these
clusters there must be enough dark matter present to stabilize
them.  This was, originally, one of the first pieces of evidence
for the existence of dark matter.

The other argument was concerned with the ages of the galaxies.
Until fairly recently it has been argued that all galaxies have
stellar populations which include stars which are very old, with
ages on the order of $H_o^{-1}$, i.e. that they are all as old as
the classic big bang universe.  However we now know that young
galaxies with ages $\ll H_o^{-1}$ do exist.  But the major point
made by Ambartsumian was, and is, that there are large numbers of
clusters of galaxies, and many small groups, which are physically
connected but clearly from their forms and their relative
velocities, appear to be unstable.

In this situation the use of the virial theorem is totally
inappropriate. It is worthwhile pointing out that if the virial
theorem holds the random motions of the galaxies should follow a
steady state distribution such as

\begin{equation}
F({\bf v}) \propto {\rm exp}\left[- {{\bf v}^2 \over 2\sigma^2}\right] .
\end{equation}

So far there is no observational demonstration that this is indeed the
case.  The conclusion drawn from  $2E_K + \Omega > O$ as based on
visible components only should rather be that the clusters are manifestly
$not$ in dynamical equilibrium.

Unfortunately, over the last thirty years the virial approach has
been wedded to the idea that all galaxies are old, and it is this
mis-reading of the data that led to the view that most galaxies
were formed in the early universe and cannot be forming now.  For
example, in 1974 Ostriker, Peebles and Yahil (1974) argued in a
very influential paper that the masses of physical systems of
galaxies increase linearly with their sizes.  As one of us pointed
out at the time (Burbidge, 1975) this result was obtained
completely by assuming that at every scale, for binary galaxies,
very small groups, larger groups, and rich clusters, the virial
condition of stability holds. Thus it was argued that more and
more dark matter is present as the systems get bigger.

Modern evidence concerning the masses of clusters has been
obtained from x-ray studies, the Sunyaev-Zeldovich effect, and
gravitational lensing (cf. Fabian 1994; Carlstrom et al. 2002;
Fort and Mellier 1994 and many other papers).  All of these
studies of rich clusters of galaxies show that large amounts of
matter in the form of hot gas and/or dark matter must be present.
However, evidence of enough matter to bind small or irregular
clusters has not been found in general, and these are the types of
configurations which Ambartsumian was originally considering. A
system such as the Hercules Cluster is in this category. Also the
very compact groups of galaxies (cf. Hickson 1997) have been a
subject of debate for many years since a significant fraction of
them ($\sim$ 40\%) contain one galaxy with a redshift very
different from the others. Many statistical studies of these have
been made, the orthodox view being that such galaxies must
be``interlopers"; foreground or background galaxies. Otherwise
they either have anomalous redshifts, or are exploding away from
the other galaxies.

We also have the problem of interacting galaxies, briefly referred
to earlier in Section 1.  In modern times it has been generally
supposed that when two galaxies are clearly in interaction they
must be coming together (merging) and never coming apart.  There
are valid ways of deciding whether or not mergers are, or have
occurred.  The clearest way to show that they are coming together
is to look for tidal tails (Toomre and Toomre 1972), or, if they
are very closely interwoven, to look for two centers, or two
counter rotating systems.  For some objects this evidence does
exist, and mergers are well established.  But to assume that
merging is occurring in all cases is unreasonable: there may well
be systems where we are seeing the ejection of one galaxy from
another as Ambartsumian proposed.  Thus when the virial condition
is not satisfied, and the systems are highly irregular and appear
to be coming apart, then perhaps they {\it are} coming
apart, and never have been separate.  Here we are clearly
departing from the standard point of view. 

If one assumes that clusters may not be bound, their overall astrophysics
changes from that of bound `steady' clusters.  Issues like the nature of
intracluster medium, the role of the halo, generation of x-rays will require
a new approach in the case where clusters are expanding.  Further, 
the ejection of new matter provides additional inputs to the dynamics 
of the system.  For example, the energy of ejection will play a role in 
heating the intracluster gas.  This important investigation still needs to be
carried out.  However, a preliminary discussion may be found in Hoyle, et al.
(2000), Chapter 20.

\subsection{Explosions in individual galaxies}

By the early 1960s it had become clear that very large energy
outbursts are taking place in the nuclei of galaxies.

The first evidence came from the discovery of powerful radio
sources and the realization that the nuclei of the galaxies which
they were identified with, had given rise to at least 10$^{59}$ -
10$^{61}$ ergs largely in the form of relativistic (Gev) particles
and magnetic flux which had been ejected to distances of $\geq$
100 kpc from the region of production.

A second line of evidence comes from the classical Seyfert
galaxies which have very bright star-like nuclei which show very
blue continua, and highly excited gas which has random motions
$\gtrsim$ 3000 Km sec$^{-1}$, and must be escaping from the
nucleus.  We know that the gas is being ejected because we see it
through absorption in optical and X-ray spectra of Seyfert nuclei,
and the wavelengths of the absorption lines are shifted to the
blue of the main emission.  The speeds observed are very large
compared with the escape velocity.  Early data were described by
Burbidge et al. (1963).

In the decades since then it has been shown that many active
nuclei are giving rise to x-rays, and to relativistic jets,
detected in the most detail as high frequency radio waves.  A very
large fraction of all of the energy which is detected in the
compact sources is non-thermal in origin, and is likely to be
incoherent synchrotron radiation or Compton radiation.

Early in the discussion of the origin of these very large energies
it was concluded that the only possible energy sources are
gravitational energy associated with the collapse of a large mass,
and the ejection of a small fraction of the energy, or we are
indeed seeing mass and energy being created in the nuclei (cf.
Hoyle, Fowler, Burbidge and Burbidge 1964).

Of course the most conservative explanation is that the energy
arises from matter falling into massive black holes with an
efficiency of conversion of gravitational energy to whatever is
seen, of order 10\%.  This is the argument that has been generally
advanced and widely accepted (cf. Rees 1984).

Why do we believe that this is not the correct explanation? After
all, there is good evidence that many nearby galaxies (most of
which are not active) contain collapsed supermassive objects in their 
centers with masses in the range 10$^6$ - 10$^8$ M$_\odot$.

The major difficulty is associated with the efficiency with which
gravitational energy can be converted into very fast moving gas
and relativistic particles, a problem that has haunted us for more
than forty years (Burbidge and Burbidge 1965).  In our view the
efficiency factor is not 10\% but close to 0.1\% - 1\%. The
reasons why the efficiency factor is very small are the following.
If the energy could be converted directly the efficiency might be
as high as $\sim$ 8\%, or even higher from a Kerr rotating black
hole. But this energy will appear outside the Schwarzschild radius
as the classical equivalent of gravitons. This energy has to be
used to heat an accretion disk or generate a corona in a classical
AGN, or generate very high energy
particles which can propagate outward in a radio source, then heat
gas which gives rise to shock waves, which accelerate particles,
which in turn radiate by the synchrotron process.  Thermodynamics
tells us that the efficiency at each of these stages is $\lesssim$
10\%. If there are 3 to 4 stages the overall efficiency is $\sim$
10$^{-3}$ - 10$^{-4}$. This is borne out by the measured
efficiency by which relativistic beams are generated in particle
accelerators on earth, and by the efficiency associated with the
activity in the center of M87. (cf. Churasov et al. 2002).

If these arguments are not accepted, and gravitational energy is
still claimed to be the only reasonable source, another problem
appears.

For the most luminous sources, powerful radio sources and distant
QSOs the masses involved must be much greater than the typical
values used by the black hole-accretion disk theorists.  If one uses the 
formula for Eddington luminosity (cf. for details pages 109-111, 408-409 
of Kembhavi \& Narlikar 1999) one arrives at black hole masses of the 
order $10^8$ M$_\odot$ on the basis of perfect efficiency of energy 
conversion.  An efficiency of $\leq 0.01$ would drive the mass up a 
hundred fold at least, i.e. to 10$^{10}$ M$_\odot$ or greater.  So far there is no direct
evidence in any galaxy for such large dark masses.  The largest
masses which have been reliably estimated are about 10$^{9}$
M$_\odot$.

In general it is necessary to explain where the bulk of the energy
released which is not in the relativistic particle beams, is to be
found. A possible explanation is that it is much of this energy
which heats the diffuse gas in active galaxies giving rise to the
extended X-ray emission in clusters and galaxies.

An even harder problem is to explain how the massive black holes
in galaxies were formed in the first place.  Were they formed
before the galaxies or later? In the standard model both scenarios
have been tried, but no satisfactory answer has been found.

In our model the energy comes with creation in the very strong
gravitational fields very close to the central NBH, where the
process can be much more efficient than can be expected in the
tortuous chain envisaged in the classical gravitational picture.
We shall discuss this in Section 7. 

Would very massive galaxies result if the universe allows indefinitely 
large time for galaxy formation?  Earlier ideas (Hoyle, 1953, Binney 1977, 
Rees and Ostriker 1977, Silk 1977) seemed to suggest so.  In the present
case two effects intervene to make massive galaxies rather rare.  The first
one is geometrical.  Because of steady long-term expansion, the distance 
between two galaxies formed, say, $n$ cycles ago, would have increased
by a factor $\sim {\rm ~exp~} n~Q/P$, and their density decreased by the
factor $\sim {\rm ~exp} -3n Q/P$. For $n \gg 1$, we expect the chance
of finding such galaxies very small.  

The second reason working against the growth of mass in a region comes
from the negative energy and pressure of the $C$-field.  As the mass 
grows through creation, the $C$-field also mounts and its repulsive 
effect ultimately causes enough instability for the mass to break up.  Thus
the large mass grows smaller by ejecting its broken parts.  

What is ejected in an MCE?  Are the ejecta more in the form of particles 
or radiation or coherent objects?  All three are produced.  For a discussion
of the mechanism leading to ejection of coherent objects, see Hoyle, et al. (2000),
Chapter 18.     

\subsection{Quasi-Stellar Objects}

In the early 1960s QSOs were discovered (Matthews and Sandage 1963; 
Schmidt 1963; cf. Burbidge and Burbidge 1967 for an extensive discussion) as 
star-like objects with large redshifts. Very early on, continuity arguments
led to the general conclusion that they are very similar to the
classical Seyfert glaxies, i.e. they are the nuclei of galaxies at
much greater distances. However, also quite early in the
investigations, it became clear that a good case could also be
made for supposing that they are more likely to be compact objects
$ejected$ from comparatively local, low redshift active galaxies
(Hoyle and Burbidge 1966).

After more than thirty years of controversy this issue has not yet
been settled, but a very strong case for this latter hypothesis
based on the observations of the clustering of many QSOs about
active galaxies has been made. (Burbidge et al. 1971; Arp 1987;
Burbidge 1996).

If this is accepted, it provides direct evidence that in the
creation process active galaxies are able to eject compact sources
with large intrinsic redshifts.  What was not predicted was the
existence of intrinsic redshifts.  They present us with an
unsolved problem, but one which must be closely connected to the
creation process. A remarkable aspect of this problem is that the
intrinsic redshifts show very clear peaks in their distribution
with the first peak at $z = 0.061$ and with a periodicity of the
form $\triangle$ log (1 + $z$) = 0.089 (cf. Karlsson 1971, Burbidge
and Napier 2001).  The periodicity is in the intrinsic redshift
component ($z_i$), and in order to single out that component,
either the cosmological redshift component $z_c$ must be very
small i.e., the sources must be very close to us, or it must be
known and corrected for by using the relation $(1 + z_{obs})$ = (1
+ $z_c$)(1 + $z_i$). Thus a recent claim that the periodicity is
not confirmed (Hawkins et al., 2003) has been shown to be in error
(Napier and Burbidge, 2003).  

It is admitted that the evidence from gravitational lensing provides an 
overall consistent picture for the standard cosmological hypothesis.  The
evidence on quasars of larger redshift being lensed by a galaxy of lower
redshift, together with the time delay in the radiation found in the two
lensed images can be explained by this hypothesis.  This type of 
evidence needs to be looked at afresh if the claim is made that quasars
are much closer than their redshift-distances.  In such cases, the lensing
models can be `scaled' down but the time-delay will have to be checked
for lower values.  To our knowledge no such exercise has been carried
out to date. We hope to examine this issue in a later paper.

\subsection{Gamma Ray Bursts}

One of the most remarkable phenomena discovered in recent years
relate to very short lived ($\lesssim$ minutes) bursts of high
energy photons ($\gamma$-ray and x-ray) which can
apparently occur anywhere in the sky, and which sometimes can be
identified with a very faint optical and/or radio source, an
afterglow, which may fade with time.  Sometimes a very faint
object remains.  The first optical observation in which a redshift
could be measured led to the conclusion that those sources are
extragalactic.  Using the redshifts as distance indicators this
has led to the conclusion that the energies emitted lie in the
range 10$^{50}$ - 10$^{54}$ ergs, with most of them $\gtrsim$
10$^{53}$ ergs, if the explosions take place isotropically.  If
energies involving single stars are invoked the energies can be
reduced if beaming is present.  The most recent observations have
suggested that the events are due to forms of supernovae which are
beamed.  In the usual interpretation it is assumed that the
redshifts which have been measured for the gamma ray bursts are
cosmological (cf Bloom et al. 2003).  However in a recent study
using all (more than 30) gamma-ray bursts (GRBs) with measured
redshifts it was shown that the redshift distribution strongly
suggests that they are closely related to QSOs with the same
intrinsic redshift peaks (Burbidge 2003, 2004). Also an analysis of
the positions of all of the GRBs for which we have positions
(about 150) shows that a number of them are very near to already
identified QSOs (Burbidge 2003). All of this suggests that the
GRBs are due to explosions of objects (perhaps $in$ QSOs) which
have themselves been ejected following a creation process from
active galaxies. In general they have slightly greater
cosmological redshifts and thus are further away ($\leq$ 500 Mpc)
than the galaxies from which most of bright QSOs are ejected.
While we do not claim that this hypothesis is generally accepted,
Bloom (2003) has shown that there are peculiarities in the
redshift distribution interpreted in the conventional way. More
observations may clarify this situation.

\section{Dynamics and Spectrum of Radiation from a
MCE}

A discussion of how a minicreation event arises was given in section 3.  Thus
we took the modified problem of a collapsing dust ball
in the presence of the $C$-field as a toy-model of how a realistic massive 
object would behave.  In the classic Oppenheimer-Snyder case the dust ball
collapses to become a black hole, eventually ending in spacetime singularity.  In
the modified problem, as we saw in section 3, the dust ball need not become a
black hole.  It certainly does not attain singularity, but bounces at a finite radius.
We saw that after bounce its outward speed rapidly rises before it ultimately 
slows down to a halt.  In the phase of rapid expansion it resembles the classical white hole -
which is the reverse of the classical collapse without the $C$-field.  The white
hole solution can be used to approximate the behaviour of an MCE as seen by
an external observer, because the former can be handled exactly in analytic 
way.  In essence we use the notation of section 3 with slight modification.  

We begin with a discussion of a white hole as considered by
Narlikar et al (1974) within the framework of standard general
relativity.  Consider a massive object emerging from a spacetime
singularity in the form of an explosion.  To simplify matters
Narlikar, Apparao and Dadhich (op. cit.) considered the object as
a homogeneous dust ball, for which one can use comoving
coordinates.  As described in section 3, the line element within
the object is given by

\begin{equation}
{\rm d}s^{2} = {\rm d}t^{2}-a^{2}(t)\left[{{\rm d}r^2\over (1- \alpha
r^{2})}+r^{2}({\rm d} \theta ^{2}+{\rm ~sin}^{2} \theta~ {\rm d}
\phi^{2})\right]
\end{equation}

\noindent where $c$ = speed of light is  taken as unity, $a(t)$ is the expansion
factor and $\alpha$ is a parameter related to the mass $M$ and the
comoving radius $r_b$ of the object by

\begin{equation}
2GM = \alpha r_{b}^{3}
\end{equation}

The similarity of equation (48) to the Robertson-Walker line
element of cosmology is well known. Also, if we change $t$ to
$-t$, equation (48) represents a freely collapsing ball of dust.
The parameter $\alpha$ is related to the dust density $\rho_0$ at
$a = 1$, by the relation

\begin{equation}
\alpha = \frac{8\pi G\rho_0}{3}.
\end{equation}

The formulae (48) - (50) are the same as (5), (10) and (6) of section 3.  
However, in $\S$ 3 we were discussing the contracting phase, while here we
are interested in the expanding mode. For convenience  therefore, we will measure $t$ from 
the instant of explosion so that $a(0) = 0$.  For $t > 0$, $a(t)$ satisfies the equation

\begin{equation}
\dot{a}^{2}= {\alpha (1-a) \over a},
\end{equation}

\noindent so that it attains its maximum value $a=1$ at

\begin{equation}
t = t_{0} = {\pi \over (2\surd \alpha)}.
\end{equation}

\noindent We will investigate light emission from the white hole
in the interval $0<t<t_{0}$. The equation (51) can be solved in a
parametric form by defining

\begin{equation}
a= {~\rm sin}^{2}\xi, ~~0 \leq \xi \leq \pi /2.
\end{equation}

The $\xi$ is related to the comoving time coordinate $t$  by

\begin{equation}
t = {2t_{0} \over \pi} (\xi - {~\rm sin}~ \xi ~{\rm cos}~ \xi).
\end{equation}

\noindent The white hole bursts out of the Schwarzschild radius at
$t=t_{c}$, $\xi = \xi_{c}$, where

\begin{equation}
{\rm sin}~ \xi_{c} = (\alpha r_{b}^{2})^{1/2}.
\end{equation}

The space exterior to the white hole is described by the
Schwarzschild line element

\begin{eqnarray}
{\rm d}s^{2}=[1 -(2GM/R)]{\rm d}T^{2}-{{\rm
d}R^{2} \over 1-(2GM/R)} \nonumber \\
-R^{2}({\rm d} \theta^{2}+{~\rm sin}^{2}
\theta {\rm d} \phi^{2}).
\end{eqnarray}

\noindent A typical Schwarzschild observer has $R$ = constant,
$\theta$ = constant, $\phi$ = constant.  We wish to calculate the
spectrum of radiation from the white hole as seen by a
Schwarzschild observer with $R=R_{1} \gg 2GM$. To simplify
matters further, we will take the luminosity spectrum of the white
hole as $L \delta (\nu-\nu_{0})$, where $L$ = constant.

Suppose two successive light signals are sent out from the surface
at comoving instants $t$ and $t + dt$ and are received by the
observer at $R_1$ at instants $T$ and $T+{\rm d}T$ measured in the
Schwarzschild coordinates.  Then a straightforward calculation
shows that

\begin{equation}
{{\rm d}T \over {\rm d}t} = {{\rm sin}~ \xi \over {\rm sin}(\xi +
\xi_{c})}.
\end{equation}

\noindent So an electromagnetic wave of frequency $\nu_0$ emitted
from the surface appears to the receiver to have the frequency

\begin{equation}
\nu = \nu_{0}\left[{{\rm sin}(\xi + \xi_{c}) \over {\rm sin}~\xi}\right].
\end{equation}

A result of this type is suitable for
working out the spectrum of the radiation as seen by the
Schwarzschild observer.  Under our assumption $L/h\nu_{0}$ photons
of frequency $\nu_0$ are being emitted per unit $t -$ time from
the surface.  The number emitted in the interval $[t,~t+{\rm d}t]$
is therefore $L{\rm d}t/h\nu_{0}$. The same number must be
received in the interval $[T,~T+ {\rm d}T]$, but with frequencies
in the range $(\nu,~\nu+{\rm d}\nu)$ where ${\rm d}\nu$ is related
to ${\rm d}t$ through equations (54) and (58).  A simple calculation gives

\begin{equation}
{\rm d}t = {(4t_{0}\nu_{0}^{3}{\rm ~sin}^{3}\xi_{c}{\rm ~
d}\nu) \over \pi(\nu^{2}+\nu^{2}_{0}-2\nu \nu_{0}{\rm
~cos ~}\xi_c)^{2}}.
\end{equation}

Writing $E=h\nu, ~E_{0}=h\nu_{0}$, the number of photons
in the range $[E,~E-{\rm d}E]$ received from the white hole per
unit area at $R=R_{1}$ is given by

\begin{equation}
N(E)~{\rm d}E={Lt_{0}\over\pi^{2}R_{1}^{2}}\times{E_{0}^{2}~{\rm
sin}^{3}\xi_{c}~{\rm d}E\over(E^{2}+E_{0}^{2}-2EE_{0}{\rm
~cos}~\xi_{c})^{2}}.
\end{equation}

\noindent For $E \gg E_{0}$

\begin{equation}
N(E)~{\rm d}E \cong Lt_{0}E_{0}^{2}\times{{\rm
sin}^{3}\xi_{c}\over\pi^{2}R_{1}^{2}} ~{{\rm d}E\over E^{4}}.
\end{equation}

\noindent The energy spectrum $I(E)$ is given by

\begin{equation}
I(E) = EN(E) \propto E^{-3}.
\end{equation}

This is the spectrum at the high energy end under the simplifying
assumptions made here. More general (and perhaps more realistic)
assumptions can lead to different types of spectra which can also
be worked out.  Following Narlikar et al (1974) possible fields in
high energy astrophysics where MCEs might find applications
are as follows.

\noindent (i)  The hard electromagnetic radiation from the MCEs
situated at the centres of, say Seyfert galaxies, can be a source
of background X and gamma radiation.  The energy spectrum (60)
seems, at first sight to be too steep compared to the observed
spectrum $\propto E^{-1.2}$.  But absorption effects in the gas
present in the nuclei surrounding the MCE tend to flatten
the spectrum given by equation (60). Detailed calculation with
available data shows that these absorption effects can in fact
flatten the $E^{-3}$ spectrum to $\sim E^{-1}$ form in the range
0.2 keV to 1keV.  At lower energies, the ultraviolet radiation
seems to be of the right order of magnitude to account for the
infrared emission of $\sim 10^{45} ~{\rm erg}~ {\rm s}^{-1}$
through the dust grain heating mechanism.

\noindent (ii)  The transient nature of X-ray and gamma-ray bursts
suggests an MCE origin.  The shape of the spectrum at the
emitting end is likely to be more complicated than the very simple
form assumed in the above example.  {\it In general, however, the
spectrum should soften with time.}

\noindent (iii)  Although Narlikar et al. (1974) had worked out the
spectrum of photons, it is not difficult to see that similar
conclusions will apply to particles of non-zero rest mass provided
they have very high energy, with the relativistic $\gamma-$ factor
$\gg 1$.  It is possible therefore to think of MCEs in the
Galaxy on the scale of supernovae, yielding high energy cosmic
rays right up to the highest energy observed.

This picture of a white hole gives a quantitative but approximate description of
radiation coming out of an MCE, which is without a
singular origin and without an event horizon to emerge out of.  Ideally we
should have used the modified $C$-field solution described in section 3
to calculate the exact result.  This, however has proved to be an intractable 
problem analytically as an explicit exterior solution is not known.  

The collapse problem with an earlier version of the $C$-field was discussed by 
Hoyle and Narlikar (1964) in which a proof was given that an exterior solution 
matching the homogeneous dust ball oscillation exists.  However an explicit 
solution could not be given.  The same difficulty exists with this solution also
and further work, possibly using numerical general relativity, may be required.  
We mention in passing that a similar matching problem exists in inflationary models where 
a Friedmann bubble emerges within an external de Sitter type universe.  

The above type of expansion has one signature.  Its explosive nature
will generate strong blueshifts, thus making the radiation of high
frequency, which softens to that at lower frequencies as the
expansion slows down. This model therefore has the general
features associated with gamma ray bursts and transient X-ray
bursters.

A further generalization of this idea at a qualitative level
corresponds to the introduction of spin so as to correspond to the
Kerr solution in classical general relativity.  If we consider an MCE to
have axial symmetry because of spin, the tendency to go round the
axis is strong in a region close to the `equator' and not so
strong away from it. In classical general relativity the
ergosphere identifies such a region: it shrinks to zero at the
poles. At the poles therefore we expect that the ejection outwards will be
preferentially directed along the axis and so we may see
jets issuing in opposite directions. 

In the very first paper on the QSSC, Hoyle, et al. (1993) had pointed to the
similarity between an MCE and the standard early universe.  In particular 
they had shown that the creation of matter in the form of Planck particles
leads to their subsequent decay into baryons together with release of very
high energy.  These `Planck fireballs' have a density temperature relationship
of the form $\rho \alpha T^3$ which permits the synthesis of light nuclei 
just as in the classical big bang model.  However, these authors drew attention 
to the circumstance that the relevant $(\rho, T)$ domain for this purpose 
in the QSSC is very different from the $(\rho, T)$ domain in the primordial 
nucleosynthesis of standard cosmology.

\section{Concluding Remarks}

The oscillating universe in the QSSC, together with a long-term
expansion, driven by a population of mini-creation events provides
the missing dynamical connection between cosmology and the `local'
explosive phenomena. The QSSC additionally fulfills the roles
normally expected of a cosmological theory, namely (i) it provides
an explanation of the cosmic microwave background with
temperature, spectrum and inhomogeneities related to astrophysical
processes (Narlikar et al. 2003), (ii) it offers a purely
stellar-based interpretation of all observed nuclei ({\it
including} light ones)(Burbidge et al. 1957; Burbidge and Hoyle
1998), (iii) it generates baryonic dark matter as part of stellar
evolution (Hoyle et al. 1994), (iv) it accounts for the extra
dimming of distant supernovae {\it without} having recourse to
dark energy (Narlikar, Vishwakarma and Burbidge 2002; Vishwakarma and
Narlikar 2005), and it also
suggests a possible role of MCEs in the overall scenario of
structure formation (Nayeri et al. 1999).  

The last mentioned work shows that preferential creation of new matter
near existing concentrations of mass can lead to growth of clustering.  A toy
model based on million-body simulations demonstrates this effect and leads
to clustering with a 2-point correlation function with index close to $-1.8$.  
Because of repulsive effect of the $C$-field, it is felt that this process may be
more important than gravitational clustering.  However, we need to demonstrate
this through simulations like those in our toy model, {\it together with} gravitational
clustering.  

There are two challenges that still remain, namely understanding
the $origin$ of anomalous redshifts and the observed
$periodicities$ in the redshifts.  Given the QSSC
framework, one needs to find a scenario in which the hitherto
classical interpretation of redshifts is enriched further with
inputs of quantum theory.  These are huge problems which we
continue to wrestle with.

\bigskip

\noindent {\it Acknowledgements}

One of us, (JVN) thanks College de France, Paris for hospitality when
this work was in process. RGV is grateful to IUCAA for hospitality which
facilitated this collaboration.

\clearpage

\section* {References}

\noindent Ambartsumian, V.A. 1965, {\it Structure and Evolution of
Galaxies, Proc. 13th Solvay
Conf. on Physics, University of Brussels}, (New York, Wiley Interscience), 24 \\

\noindent Arp, H.C. 1987, {\it Quasars, Redshifts and
Controversies} (Interstellar Media, Berkeley, California) \\ 

\noindent Bagla, J.S., Padmanabhan, T. and Narlikar, J.V. 1996, {\it Comm. Astrophys.}, 
{\bf 18}, 289 \\

\noindent Barrow, J., Kimberly, D. and Magueijo, J. 2004, {\it Class. Quant. Grav.}, {\bf 21},
4289 \\

\noindent Binney, J. 1977, {\it Ap.J.}, {\bf 215}, 483 \\

\noindent Blanchard, A., Souspis, B., Rowan-Robinson, M. and
Sarkar, S. 2003, {\it A\&A}, {\bf 412}, 35 \\

\noindent Bloom, J.S. 2003, {\it A.J.}, {\bf 125}, 2865 \\

\noindent Bloom, J.S., Kulkarni, S.R. and Djorgovsky, S.G. 2001, {\it A.J.}, {\bf 123}, 1111 \\ 

\noindent Bondi, H. and Gold, T. 1948, {\it MNRAS}, {\bf 108}, 252 \\

\noindent Burbidge, E.M., Burbidge, G.R., Fowler, W.A. and Hoyle, F. 1957, {\it Rev. Mod. Phys.}, {\bf 29}, 547 \\

\noindent Burbidge, E.M., Burbidge, G., Solomon, P. and
Strittmatter, P.A. 1971, {\it Ap.J.}, {\bf 170}, 223 \\

\noindent Burbidge, G. 1975, {\it Ap.J.}, {\bf 106}, L7 \\

\noindent Burbidge, G. 1996, {\it A\&A}, {\bf 309}, 9 \\

\noindent Burbidge, G. 2003, {\it Ap.J.}, {\bf 585}, 112\\

\noindent Burbidge, G. 2004, ``The Restless High Energy Universe'', 
{\it Conf. Proc. Nuclear Physics B.}, {\bf 305}, 132 \\

\noindent Burbidge, G. and Burbidge, E.M. 1965, {\it The Structure and Evolution of
Galaxies, Proc. of 13th Solvay Conference on Physics, University of Brussels}, (New York,
Wiley Interscience), 137 \\

\noindent Burbidge, G. and Burbidge, E.M. 1967, {\it Quasi-Stellar
Objects}, (San Francisco, W.H. Freeman) \\

\noindent Burbidge, G. and Hoyle, F. 1998, {\it ApJ.}, {\bf 509}, L1 \\

\noindent Burbidge, G. and Napier, W. M. 2001, {\it A.J.}, {\bf 121}, 21 \\

\noindent Burbidge, G., Burbidge, E.M., and Sandage, A. 1963, {\it Rev. Mod. Phys}., {\bf 35}, 947\\

\noindent Carlstrom, J., Holder, G. and Reese, E. 2002, {\it A.R.A.A.}, {\bf 40}, 643 \\ 

\noindent Carroll, S.M. and Press, W.H. 1992, {\it A.R.A.A.}, {\bf 30}, 499 \\

\noindent Churasov, E., Sunyaev, R., Forman, W. and Bohringer, H.
2002, {\it MNRAS}, {\bf 332}, 729 \\

\noindent Datt, B. 1938, {\it Z. Phys.}, {\bf 108}, 314 \\

\noindent Fabian, A.C. 1994, {\it A.R.A.A.}, {\bf 32}, 277 \\

\noindent Fort, B. and Mellier, Y. 1994, {\it A\&A Rev.}, {\bf 4}, 239 \\

\noindent Gliner, E.B. 1970, {\it Soviet Physics-Doklady}, {\bf 15}, 559 \\

\noindent Gunn, J.B. and Oke, J.B. 1975, {\it Ap.J.}, {\bf 195}, 255\\

\noindent Hawking, S.W. and Ellis, G.F.R. 1973, {\it The Large Scale Structure 
of Space-time}, Cambridge \\

\noindent Hawkins, E., Maddox, S.J. and Merrifield, M.R. 2002,
{\it MNRAS}, {\bf 336}, L13 \\

\noindent Hickson, P. 1997, {\it A.R.A.A.}, {\bf 35}, 377 \\

\noindent Hogarth, J.E. 1962, {\it Proc. R. Soc.}, {\bf A267}, 365 \\ 

\noindent Hoyle, F. 1948, {\it MNRAS}, {\bf 108}, 372 \\

\noindent Hoyle, F.1953, {\it Ap.J}, {\bf 118}, 513 \\

\noindent Hoyle, F. and Burbidge, G. 1966, {\it Ap.J.}, {\bf 144}, 534 \\

\noindent Hoyle, F. and Narlikar, J.V 1963, {\it Proc. Roy. Soc.}, {\bf A277}, 1 \\

\noindent Hoyle, F. and Narlikar, J.V. 1964, {\it Proc. Roy. Soc.}, {\bf A278}, 465 \\

\noindent Hoyle, F. and Narlikar, J.V. 1969, {\it Ann. Phys. (N.Y.)}, {\bf 54}, 207 \\

\noindent Hoyle, F. and Narlikar, J.V. 1971, {\it Ann. Phys. (N.Y.)}, {\bf 62}, 44 \\

\noindent Hoyle, F. and Narlikar, J.V. 1995, {\it Rev. Mod. Phys.}, {\bf 61}, 113 \\

\noindent Hoyle, F., Burbidge, G. and Narlikar, J.V. 1993, {\it Ap.J}, {\bf 410}, 437 \\

\noindent Hoyle, F., Burbidge, G. and Narlikar, J.V. 1994, {\it MNRAS}, {\bf 267}, 1007 \\

\noindent Hoyle, F., Burbidge, G. and Narlikar, J.V. 1995, {\it Proc. Roy. Soc.}, {\bf A448}, 191 \\

\noindent Hoyle, F., Burbidge, G. and Narlikar, J.V. 2000, {\it A Different Approach to
Cosmology}, (Cambridge, Cambridge University Press). \\

\noindent Hoyle, F., Fowler, W.A., Burbidge, E.M. and Burbidge, G. 1964, {\it Ap.J},
 {\bf 139}, 909 \\

\noindent Hoyle, F. and Sandage, A. 1956, {\it P.A.S.P.}, {\bf 68}, 301 \\

\noindent Karlsson, K.G. 1971, {\it A\&A}, {\bf 13}, 333 \\  

\noindent Kembhavi, A.K. and Narlikar, J.V. 1999, {\it Quasars and Active Galactic 
Nuclei}, (Cambridge, Cambridge University Press). \\

\noindent Longair, M.S. 1987, {\it IAU Symposium 124,
``Observational Cosmology"}, (Editors,  A. Hewitt, G. Burbidge,
L.Z. Fang:  D. Reidel, Dordrecht) p. 823 \\

\noindent Matthews, T.A. and Sandage, A.R. 1963, {\it Ap.J.}, {\bf 138}, 30 \\  

\noindent McCrea, W.H. 1951, {\it Proc.Roy.Soc.}, {\bf A206}, 562 \\

\noindent Meyers, A.D., Shanks, T., Outram, J.J., Srith, W.J. and
Wolfendale, A.W. 2004, {\it MNRAS}, {\bf 347}, L67 \\

\noindent Napier, W. and Burbidge, G. 2003, {\it MNRAS}, {\bf
342}, 601 \\

\noindent Narlikar, J.V. 1973, {\it Nature}, {\bf 242}, 35 \\

\noindent Narlikar, J.V. and Padmanabhan, T. 1985, {\it Phys. Rev.} {\bf D32}, 1928 \\

\noindent Narlikar, J.V., Apparao, M.V.K. and Dadhich, N.K. 1974, {\it Nature}, {\bf 251}, 590 \\

\noindent Narlikar, J.V., Vishwakarma, R.G. and Burbidge, G.  2002, {\it P.A.S.P.}, {\bf 114}, 1092 \\

\noindent Narlikar, J.V., Vishwakarma, R.G., Hajian, A., Souradeep, T., Burbidge, G. and Hoyle, F. 2003, {\it Ap.J.}, {\bf 585}, 1 \\

\noindent Nayeri, A., Engineer, S., Narlikar, J.V. and Hoyle, F. 1999, {\it Ap.J.}, {\bf 525}, 10 \\

\noindent Ostriker, J.P., Peebles, P.J.E. and Yahil, A. 1974, {\it Ap.J.}, {\bf 193}, L1 \\

\noindent Page L., et al., 2003, Astrophys. J. Suppl. {\bf 148}, 233\\

\noindent Perlmutter, S. et al. 1999, {\it Ap.J.}, {\bf 517}, 565 \\

\noindent Podariu, S., Souradeep, T.,   Gott III, J. R., 
   Ratra, B.  and Vogeley, M. S. 2001, {\it Ap. J. S.}, {\bf 559}, 9\\

\noindent Rees, M.J. 1984, {\it A.R.A.A.}, {\bf 22}, 471 \\ 

\noindent Rees, M.J. and Ostriker, J.P. 1977, {\it MNRAS}, {\bf 179}, 541 \\

\noindent Riess, A. et al. 1998, {\it A.J.}, {\bf 116}, 1009 \\

\noindent Rubano, C. and Seudellaro, P. 2004, astro-ph / 0410260 \\

\noindent Sachs, R., Narlikar, J.V. and Hoyle, F. 1996, {\it A\&A}, {\bf 313}, 703 \\ 

\noindent Sami, M. and Toporensky, A. 2004, {\it Mod. Phys. Lett. A}, {\bf 19}, 1509 \\

\noindent Schmidt, M. 1963, {\it Nature}, {\bf 197}, 1040 \\ 

\noindent Silk, J. 1977, {\it Ap.J}, {\bf 211}, 638\\

\noindent Singh, P., Sami, M. and Dadhich, N. 2003, {\it Phys. Rev.}, {\bf D68}, 023522 \\ 

\noindent Spergel, D. et al. 2003, {\it Ap.J.S.}, {\bf 148}, 175 \\

\noindent Spergel, D.N., et al., 2006, astro-ph/0603449 \\

\noindent Steinhardt, P.J. and Turok, N. 2002, {\it Science}, {\bf 296}, 1436 \\

\noindent Toomre, A. and Toomre, J. 1972, {\it Ap.J.}, {\bf 178}, 623 \\

\noindent Vishwakarma, R.G. and Narlikar, J.V. 2005, {\it Int.J.Mod.Phys.D}, {\bf 14}, 2, 345 \\

\noindent Wheeler, J.A. and Feynman, R.P. 1945, {\it Rev. Mod. Phys.}, {\bf 17}, 157 \\

\noindent Wheeler, J.A. and Feynman, R.P. 1949, {\it Rev. Mod. Phys.}, {\bf 21}, 425 \\

\noindent Wickramasinghe, N.C. 2005, {\it Current Issues in Cosmology}, Proceedings of the Colloquium on `Cosmology: Facts and Problems ', Paris. (Cambridge, Cambridge University 
Press), 152 \\


\clearpage

\centerline{{\bf Appendix} : {\it Field Theory Underlying the
QSSC}}

Following Mach's principle, we begin with the hypothesis that
inertia of any particle of matter owes its origin to the existence
of all other particles of matter in the universe.  If the
particles are labelled $a, b, c, ...$ and the element of proper
time of $a^{th}$ particle in Riemannian spacetime is denoted by
$ds_a$, then we express the inertia of particle $a$ by the sum

$$M_a(A) = \sum_{b \neq a} \int \lambda_b \tilde{G} (A, B) ds_b = \sum_{b \neq a} M^{(b)} (A).$$

\noindent \hfill (A1)

\noindent where $A$ is a typical point on the world line of
particle $a$ .  $\tilde{G} (A, B)$ is a scalar propagator
communicating the inertial effect from $B$ to $A$.  The coupling
constant $\lambda_b$ denotes the intensity of the effect and
without loss of generality may be set equal to unity.  Likewise we
may replace $M_a (A)$ by a scalar mass function $M(X)$ of a
general spacetime point $X$, denoting the mass acquired by a
particle at that point.  As in Riemannian geometry we will denote
by $R_{ik}$ the Ricci tensor and by $R$ the scalar curvature.

The individual contributors to $M(X)$ are the scalar functions
$M^{(b)}(X)$, which are determined by the propagators
$\tilde{G}(X, B)$.  The simplest theory results from choice of a
conformally invariant wave equation for  $M^{(b)}(X)$,

$$\Box  M^{(b)} (X) + {1\over 6} R M^{(b)} (X) + M^{(b)}(X)^3 =
\int {\delta_4 (X, B) \over \sqrt{-g (B)}}ds_b.$$

\noindent \hfill (A2)

\noindent The expression on the right hand side identifies the
worldline of $b$ as the source.  Why conformal invariance?  In a
theory of long range interactions influences travel along light
cones and light cones are entities which are globally invariant
under a conformal transformation.  Thus a theory which picks out
light cones for global communication is naturally expected to be
conformally invariant.  (A comparison may be made with special
relativity.  The $local$ invariance of speed of light for all
moving observers leads to the requirement of local Lorentz
invariance of a physical theory.)

Although the above equation is non-linear, a simplification
results in the smooth fluid approximation describing a universe
containing a larger number of particles.  Thus $M(X) =
\sum\limits_b M^{(b)}(X)$ satisfies an equation

$$\Box  M + {1\over 6} R M + \Lambda M^3 = \sum\limits_b
\int {\delta_4 (X, B) \over \sqrt{-g (B)}}d^4s_b. $$

\noindent \hfill (A3)

\noindent What is $\Lambda$?  Assuming that there are $N$
contributing particles in a cosmological horizon size sphere, we
will get

$$\Lambda \approx N^{-2},$$

\noindent \hfill (A4)

\noindent since adding $N$ equations of the kind (A2) leads to the
cube term having a reduced coefficient by this factor, because of
the absence of cross products $M^{(b)}M^{(c)}$ type $(b \neq c)$.
Typically the observable mass in the universe is $\sim 10^{22}
M_\odot$ within such a sphere, giving $N \sim 2 \times 10^{60}$ if
the mass is typically that of a planck particle.  We shall return
to this aspect shortly.  With this value for $N$, we have

$$\Lambda \approx 2.5 \times 10^{-121}. $$

\noindent \hfill (A5)

With these definitions we now introduce the action principle from
which the field equations can be derived.  In particle-particle
interaction form it is simply

$${\cal A} = - \sum\limits_a \int M_a (A) ds_a. $$

\noindent \hfill (A6)

\noindent Expressed in terms of a scalar field function $M(X)$, it
becomes

\begin{eqnarray}
{\cal A} &= &-{1\over 2} {\displaystyle \int} (M_i M^i - {1\over
6} RM^2) \sqrt{-g} ~d^4 x
+ {1 \over 4} \Lambda {\displaystyle \int} M^4 \sqrt{-g} d^4 x \nonumber \\
& - & \sum\limits_a  \int {\delta_4 (X,A)\over \sqrt{-g (A)}} M
(X) ds_a. \nonumber
\end {eqnarray}

\noindent \hfill (A7) \vspace{8pt}

\noindent For example, the variation $M \rightarrow M + \delta M$
leads to the wave equation (A2).  The variation of spacetime
metric gives rise to gravitational equations. The variation of
particle world lines gives rise to another scalar field, however,
if we assume the worldlines to have finite beginnings.  This is
where creation of matter explicitly enters the picture.  The
characteristic mass of a typical particle that can be constructed
in the theory using the available fundamental constants $c, G $
and $\hbar$ is the Planck mass

$$m_P = \bigg({3\hbar c \over 4\pi G}\bigg)^{1/2}.$$

\noindent \hfill (A8)

\noindent We shall assume therefore that the typical basic
particle created is the Planck particle with the above mass.  We
shall take $\hbar = 1$ in what follows.  Imagine now the worldline
of such a particle beginning at a world-point $A_0$.

A typical Planck particle $a$ exists from $A_0$ to $A_0 + \delta
A_0$, in the neighborhood of which it decays into $n$ stable
secondaries, $n \simeq 6.10^{18}$, denoted by $a_1, a_2, \ldots
a_n$. Each such secondary contributes a mass field $m^{(a_r)}
(X)$, say, which is the fundamental solution of the wave equation


$$\Box  m^{(a_r)} + {1\over 6} R m^{(a_r)} + n^2 m^{(a_r)^3} =
{1\over n} \int_{A_0 + \delta A_0} {\delta_4 (X,A) \over \sqrt{-g
(A)}} da,$$

\noindent \hfill (A9)

\noindent  while the brief existence of $a$ contributes $c^{(a)}
(X)$, say, which satisfies

$$\Box  c^{(a)} + {1\over 6} R c^{(a)} + c^{(a)^3} =
\int^{A_0 + \delta A_0}_{A_0} {\delta_4 (X,A) \over \sqrt{-g (A)}}
da, $$

\noindent \hfill (A10)

\noindent  Summing $c^{(a)}$ with respect to $a, b, \ldots$ gives

$$c (X) = \sum\limits_a c^{(a)} (X), $$

\noindent \hfill (A11)

\noindent  the contribution to the total mass $M (X)$ from the
Planck particles during their brief existence, while

$$\sum\limits_a \sum\limits^n_{r = 1} m^{(a_r)} (X)= m (X) $$

\noindent \hfill (A12)

\noindent  gives the contribution of the stable secondary
particles.

Although $c(X)$ makes a contribution to the total mass function

$$M (X) = c(X) + m (X)$$

\noindent \hfill (A13)

\noindent  that is generally small compared to $M (X)$, there is
the difference that, whereas $m (X)$ is an essentially smooth
field, $c (X)$ contains small exceedingly rapid fluctuations and
so can contribute significantly to the derivatives of $c (X)$. The
contribution to $c (X)$ from Planck particles $a$, for example, is
largely contained between two light cones, one from $A_0$, the
other from $A_0 + \delta A_0$. Along a timelike line cutting these
two cones the contribution to $c (X)$ rises from zero as the line
crosses the light cone from $A_0$, attains some maximum value and
then falls back effectively to zero as the line crosses the second
light cone from $A_0 + \delta A_0$. The time derivative of
$c^{(a)} (X)$ therefore involves the reciprocal of the time
difference between the two light cones. This reciprocal cancels
the short duration of the source term on the right-hand side of
(A10). The factor in question is of the order of the decay time
$\tau$ of the Planck particles, $\sim 10^{-43}$ seconds. No matter
how small $\tau$ may be, the reduction in the source strength of
$c^{(a)} (X)$ is recovered in the derivatives of $c^{(a)} (X)$,
which therefore cannot be omitted from the gravitational
equations.

The derivatives of $c^{(a)} (X), c^{(b)} (X), \ldots$ can as well
be negative as positive, so that in averaging many Planck
particles, linear terms in the derivatives do disappear.   It is
therefore not hard to show that after such an averaging the
gravitational equations become


\begin{eqnarray}
R_{ik} - {1\over 2} g_{ik} R -3 \Lambda m^2 g_{ik} =  {6\over
m^2} \bigg[- T_{ik}
+ {1\over 6} (g_{ik} \Box m^2 - m^{2}_{;ik}) \nonumber \\
+  (m_{i} m_{k} - {1\over 2} g_{ik} m_{l} m^{l}) +  {2\over 3}
(c_i c_k - {1\over 4} g_{ik} c_l c^l)\bigg].  \nonumber
\end{eqnarray}

\noindent \hfill (A14)

Since the same wave equation is being used for $c (X)$ as for $m
(X)$, the theory remains scale invariant. A scale change can
therefore be introduced that reduces $M (X) = m (X) + c (X)$ to a
constant, or one that reduces $m (X)$ to a constant. Only that
which reduces $m (X)$ to a constant, viz

$$\Omega = {m (X) \over m_P} $$

\noindent \hfill (A15)

\noindent  has the virtue of not introducing small very rapidly
varying ripples into the metric tensor. Although small in
amplitude such ripples produce non-negligible contributions to the
derivatives of the metric tensor, causing difficulties in the
evaluation of the Riemann tensor, and so are better avoided.
Simplifying with (A14) does not bring in this difficulty, which is
why separating of the main smooth part of $M (X)$ now proves an
advantage, with the gravitational equations simplifying to

$$8 \pi G = {6 \over m^2_P} , ~~~m_P~~{\rm a~~constant}, $$

\noindent \hfill (A16)

$$R_{ik} - {1\over 2} g_{ik} R + \lambda  g_{ik} = - 8 \pi G [T_{ik} - {2\over 3}
(c_i c_k - {1\over 4} g_{ik} c_l c^l)].$$

\noindent \hfill (A17)

We define the cosmological constant $\lambda$ by

$$\lambda = - 3 \Lambda m^2_P \approx -2 \times 10 ^{56} {\rm ~cm}^{-2}$$

\noindent \hfill (A18)

\noindent This value falls within the normally expected region of
the magnitude of the cosmological constant.  Note, however, that
its sign is negative!  This has been the consequence of the
Machian origin of the cosmological constant through the non-linear
equations (A2), (A3).

\noindent  \hfill

\noindent  It has been on (A17) that the discussion of what is called the quasi-steady state cosmological model (QSSC) has
been based. A connection with the $C$-field of the earlier steady
state cosmology  can also be given. Writing


$$C (X) = \tau c (X),$$

\noindent \hfill (A19)

\noindent  where $\tau$ is the decay lifetime of the Planck
particle, the action contributed by Planck particles $a, b,
\ldots$,

$$- \sum\limits_a \int_{A_0}^{A_0 + \delta A_0} c (A) da$$

\noindent \hfill (A20)

\noindent  can be approximated as

$$- C (A_0) - C (B_0) - \ldots,$$

\noindent \hfill (A21)

\noindent  which form corresponds to the $C$-field used in the
steady state cosmology.

Thus the equations (A17) are replaced by

$$R_{ik} - \frac{1}{2} g_{ik} R + \lambda g_{ik} = -8 \pi G\bigg[T_{ik} - f \bigg(C_{i} C_{k} -\frac{1}{4}g_{ik}C_{l}C^{l}\bigg)\bigg],$$

\noindent \hfill (A22)

\noindent with the earlier coupling constant $f$ defined as

$$f=\frac{2}{3\tau^2}$$

\noindent \hfill (A23)

\noindent [We remind the reader that we have taken the speed of
light $c=1.$]

The question now arises of why astrophysical observation suggests
that the creation of matter occurs in some places but not in
others. For creation to occur at the points $A_0, B_0, \ldots$ it
is necessary classically that the action should not vary with
respect to small changes in the spacetime positions of these
points, which was shown earlier to require

$$C_i (A_0) C^i (A_0) = C_i (B_0) C^i (B_0) = \ldots = m_P^2.$$

\noindent \hfill (A24)

\noindent  More precisely, the field $c (X)$ is required to be
equal to $m_P$ at $A_0, B_0, \ldots ,$

$$c (A_0) = c (B_0) = \ldots = m_P.$$

\noindent \hfill (A25)

\noindent (For, equation (A19) tells us that connection between
$c$ and $C$ is through the lifetime $\tau$ of  Planck particle.)

As already remarked in the main text, this is in general not the
case: in general the magnitude of $C^iC_i$ is much less that
$m_P$. However, close to the event horizon of a massive compact
body $C_i (A_0) C^i (A_0)$ is increased by a relativistic time
dilatation factor, whereas $m_P^2$ stays fixed. Hence, near enough
to an event horizon the required conservation conditions can be
satisfied, which has the consequence that creation events occur
only in compact regions, agreeing closely with the condensed
regions of high excitation observed so widely in astrophysics.

\centerline{--------------------------------}

\label{lastpage}

\end{document}